\documentclass[prd,preprint,aps,amsmath,amssymb,showpacs,superscriptaddress,nofootinbib]{revtex4-2}

\usepackage{graphicx} 
\usepackage{dcolumn}
\usepackage{bm}
\usepackage{color}
\usepackage[normalem]{ulem}
\usepackage{comment}
\usepackage{diagbox}
\usepackage{float}
\usepackage{hyperref}

\usepackage{ulem}

\begin{document}

\preprint{\hbox{PITT-PACC-2512}}

\title{$i$-incidental $N$-naturalness}

\author{Brian Batell}
\email{batell@pitt.edu}
\affiliation{Pittsburgh Particle Physics, Astrophysics, and Cosmology Center, \\ Department of Physics and Astronomy, University of Pittsburgh, Pittsburgh, USA}
\author{Akshay Ghalsasi}
\email{aghalsasi@fas.harvard.edu}
\affiliation{Jefferson Physical Laboratory, Harvard University, Cambridge, USA}
\author{Wenjie Huang}
\email{weh68@pitt.edu}
\affiliation{Pittsburgh Particle Physics, Astrophysics, and Cosmology Center, \\ Department of Physics and Astronomy, University of Pittsburgh, Pittsburgh, USA}
\author{Matthew Low}
\email{matthew.w.low@gmail.com}
\affiliation{Pittsburgh Particle Physics, Astrophysics, and Cosmology Center, \\ Department of Physics and Astronomy, University of Pittsburgh, Pittsburgh, USA}

\date{\today}

\begin{abstract}
$N$-naturalness is a novel solution to the electroweak hierarchy problem which posits $N$ copies of 
the Standard Model with varying Higgs mass-squared parameters. 
Reheating proceeds through a ``reheaton'' particle that deposits most of its energy density into the Standard
Model and small but potentially measurable fractions into the other copies. 
Typically the sector with the lightest negative Higgs mass-squared is identified as the Standard Model. 
We demonstrate that $N$-naturalness admits a broader class of realizations in which the Standard Model is identified with a heavier sector, rather than being restricted to the lightest.
This is made possible by resonant mixing between the reheaton and the Higgs, which generically causes one sector to be preferentially reheated and to acquire the largest share of the energy density, singling it out as the Standard Model. 
We demonstrate that this scenario is consistent with current cosmological bounds on new relativistic degrees of freedom and  overclosure constraints from heavy stable relics, while future cosmic microwave background and high redshift surveys will probe significant portions of the remaining parameter space. Furthermore, we highlight the possibility of a novel stochastic gravitational wave spectrum from the many cosmological first order QCD phase transitions occurring across the other sectors.
\end{abstract}

\maketitle
\newpage

\section{Introduction}
\label{sec:introduction}

The electroweak hierarchy problem remains a central puzzle in fundamental physics and a major challenge for TeV-scale model building. 
The two primary classes of traditional solutions, supersymmetry and compositeness, both face increasingly stringent direct limits due to the steady increase of data from the Large Hadron Collider (LHC). 
 Solutions to the hierarchy problem that do not rely on compositeness or supersymmetry at low scales have become increasingly compelling~\cite{Dienes:1994np,Dienes:2001se,Dvali:2003br,Dvali:2004tma,Chacko:2005pe,Burdman:2006tz,Cheung:2014vva,Graham:2015cka,Arvanitaki:2016xds,Ibanez:2017oqr,Geller:2018xvz,Cheung:2018xnu,Craig:2019fdy,Giudice:2019iwl,Craig:2019zbn,Strumia:2020bdy,Csaki:2020zqz,Arkani-Hamed:2020yna,Giudice:2021viw,TitoDAgnolo:2021nhd,Abel:2021tyt,Csaki:2022zbc,Giudice:2017pzm,Craig:2022eqo}.  
Consequently, it is important to explore new mechanisms for explaining the weak scale and to understand their phenomenological implications.

One such theory is $N$-naturalness~\cite{Arkani-Hamed:2016rle}. In this theory there are $N$ sectors, each identical to the Standard Model (SM) except for a different Higgs mass-squared parameter $m_H^2$.
For a cut-off of $\Lambda_H$, the values of $m_H^2$ vary between $-\Lambda_H^2$ and $+\Lambda_H^2$, necessarily leading to some sectors with parametrically suppressed Higgs masses, $m_H^2 \sim \Lambda^2_H/N$.
Sectors with negative values of $m_H^2$, denoted as {\it SM-like}, have a similar particle spectrum to the SM, but heavier. 
Instead, {\it exotic} sectors with positive values of $m_H^2$ differ substantially from the SM because electroweak symmetry is broken not by the Higgs potential, but rather by the sector's QCD quark condensate. 
In the original $N$-naturalness construction~\cite{Arkani-Hamed:2016rle}, our SM is identified with the lightest SM-like sector, i.e., the sector that has a negative value of $m_H^2$ which is closest to zero.  

In this setup, the hierarchy problem is therefore recast as a cosmological question: why is our sector with a light Higgs preferentially populated in the universe?
The answer is the reheaton, a particle that dominates the energy density of the universe at early times, then distributes its energy density among the different sectors through its universal portal couplings. 
If the reheaton is light, with a mass near the weak scale, its decays to our sector are kinematically enhanced, ensuring that the SM is preferentially reheated over other sectors, as is required for a viable cosmology.

As noted above,  the original formulation of $N$-naturalness designates  the lightest sector with negative Higgs mass-squared as the SM, but this need not be assumed a priori.
The construction generically produces a small subset of sectors whose Higgs masses are light relative to the cutoff, and any one of them could in principle correspond to our SM.
The central question is whether a mechanism exists that preferentially populates one of these heavier SM-like sectors. 

In this work we show that the answer is affirmative. For a scalar singlet reheaton with universal Higgs portal couplings to the sectors, resonant reheaton–Higgs mixing  causes the reheaton to preferentially populate one of the heavier SM-like sectors. Although this might seem to require a coincidence between the reheaton mass and a particular Higgs mass, the crucial point is that such a near-degeneracy is inevitable given the closely spaced spectrum of Higgs masses across sectors. Thus, provided the reheaton is (a) heavy compared to the  Higgs of the lightest sector and (b) light compared to the cutoff, it will resonantly populate some SM-like sector with an electroweak scale far below the cutoff, singling it out as our SM, while the other sectors receive subdominant energy densities. This is a generic outcome of our construction and is in broad agreement with observational data.
 We will refer to this scenario as  $i$-incidental $N$-naturalness (since the SM resides in an incidentally selected sector with sector index $i>0$).

In the original $N$-naturalness 
as well as our $i$-incidental extension, each additional sector has a massless photon which contributes to the total relativistic energy density of the universe~\cite{Arkani-Hamed:2016rle,Choi:2018gho,Bansal:2024afn}.  Measurements of this energy density typically lead to the strongest constraints on the model parameter space.
Along with massless photons, each sector contains three neutrino species.  Depending on the mass and temperature of a neutrino species, it can contribute to the relativistic energy density, to the cold dark matter energy density, or act as warm dark matter which can suppress the matter power spectrum slightly~\cite{Bansal:2024afn} (see also \cite{Banerjee:2016suz}). 
Generically, the other sectors also contain massive stable particles that contribute a nontrivial relic abundance~\cite{Arkani-Hamed:2016rle}. We derive  constraints from cosmology on extra relativistic degrees of freedom at late times and from potential overclosure by heavy stable relics, finding that significant portions of parameter space remain open within our $i$-incidental $N$-naturalness scenario that are consistent with a solution to the little hierarchy problem, with a cutoff of $\Lambda_H \sim 10$ TeV. 
See also Refs.~\cite{Hardy:2017pnj,Baumgart:2021ptt,Easa:2022vcw,Ettengruber:2025usk} for additional studies of the cosmological implications of $N$-naturalness.

Another interesting signal that arises in $N$-naturalness models is gravitational waves~\cite{Batell:2023wdb} (see also Ref.~\cite{Archer-Smith:2019gzq}).  In the exotic sectors all 6 quarks are lighter than the corresponding QCD scale which may lead to a first-order phase transition (FOPT)~\cite{Pisarski:1983ms}.  The strength of the gravitational wave signal that results from this FOPT depends on the energy density in this sector.  It was shown in Ref.~\cite{Batell:2023wdb} that there are several regions of parameter space in the original $N$-naturalness model where the signals are potentially observable by future gravitational wave observatories. As we will show,  $i$-incidental $N$-naturalness allows for even more exotic GW spectra,  since many lighter exotic sectors can be populated (along with possible SM-like sectors containing three light quarks), each undergoing a QCD FOPT and generating a stochastic GW source. 

Our scenario has a smoking gun signature, namely, the reheaton with a mass close to the 125 GeV Higgs boson, an essential element of the reheating mechanism. In principle, this could lead to a wide variety of effects at high energy colliders that warrant experimental attention. However, a viable cosmology requires the reheating temperature to lie below the weak scale, forcing the reheaton–Higgs mixing angle to be small and rendering reheaton collider production negligible. While we mention a couple of speculative possibilities that might evade this conclusion, it appears that direct detection of the reheaton will remain challenging.

The remainder of this paper is organized as follows. In Sec.~\ref{sec:original-NN} we review the original $N$-naturalness framework and summarize its essential features. Sec.~\ref{sec:ii-NN} introduces the $i$-incidental $N$-naturalness extension, describing the reheating mechanism, properties of the various sectors, resulting cosmology, and reheaton decays. In Sec.~\ref{sec:constraints} we examine the constraints and signatures of the scenario, focusing on contributions to relativistic energy densities, overclosure from heavy relics, stochastic GW signals, and the challenges associated with probing the reheaton at colliders. We present our conclusions in Sec.~\ref{sec:outlook}. Technical details related to summing reheaton decay widths over sectors are collected in Appendix~\ref{sec:sums}.

\section{Review of $i_{\rm SM} = 0$ $N$-naturalness}
\label{sec:original-NN}

We first review the salient features of the original $N$-naturalness model~\cite{Arkani-Hamed:2016rle}. The minimal model contains $N$ copies of the SM which are mutually decoupled, with corresponding Higgs mass-squared parameters uniformly distributed, 
\begin{equation}
m_{H_i}^2 = -\frac{\Lambda_H^2}{N} (2 i +r ), ~~~~~~- \frac{N}{2} \leq i \leq  \frac{N}{2}.
\label{eq:mH-distribution-0}
\end{equation}
Here $i$ is a sector index, $\Lambda_H$ is the UV cutoff of the theory, and  $r$ controls the overall offset of the distribution ($0 \leq r \leq 2$). 
Though not strictly required, the sectors are assumed to be identical in all other respects, thus  there is a sector permutation symmetry that is softly broken by the Higgs mass parameters. 
Sectors with $i \geq 0$ are classified as {\it SM-like}, as they possess negative Higgs mass–squared parameters and  therefore undergo electroweak symmetry breaking in the standard manner.
 In contrast, sectors with $i \leq 0$ are termed {\it exotic}, as they are characterized by positive Higgs mass–squared parameters. 
In the original $N$-naturalness model, our SM is identified with the lightest SM-like sector,  $i_{\rm SM} = 0$. Therefore, we have 
$m^2_{H_0}  \equiv  \mu^2 = - (\Lambda_H^2 / N) r \approx  - (88 \, {\rm GeV})^2$, 

Beyond the presence of $N$ sectors, an essential component of $N$-naturalness is a light reheaton field, which is assumed to dominate the post-inflationary energy density of the universe for some period. 
Here we focus on the model with a real singlet scalar reheaton $\phi$, with couplings
\begin{equation}
\label{eq:Higgs-portal}
{\cal L} \supset -\frac{1}{2}m_\phi^2 \phi^2 - a \phi \sum_{i} |H_i|^2.
\end{equation}
Here $m_\phi$ is the reheaton mass and $a$ is a universal dimensionful Higgs portal coupling. Due to this coupling, the reheaton will eventually decay, and 
the fraction of reheaton energy density deposited into each sector $i$ is proportional to its partial decay width to that sector, $\Gamma_i$. To dominantly populate our SM and achieve a viable cosmology, it is crucial that the reheaton is light compared to the cutoff, with mass near the electroweak scale. Further details 
regarding reheaton decays will be presented below in Sec.~\ref{sec:reheating}. 
We now turn to the main qualitative aspects of these decays, which underlies the structure of the $N$-naturalness model.

\begin{figure}[t]
\begin{center}
\includegraphics[width=0.55\textwidth]{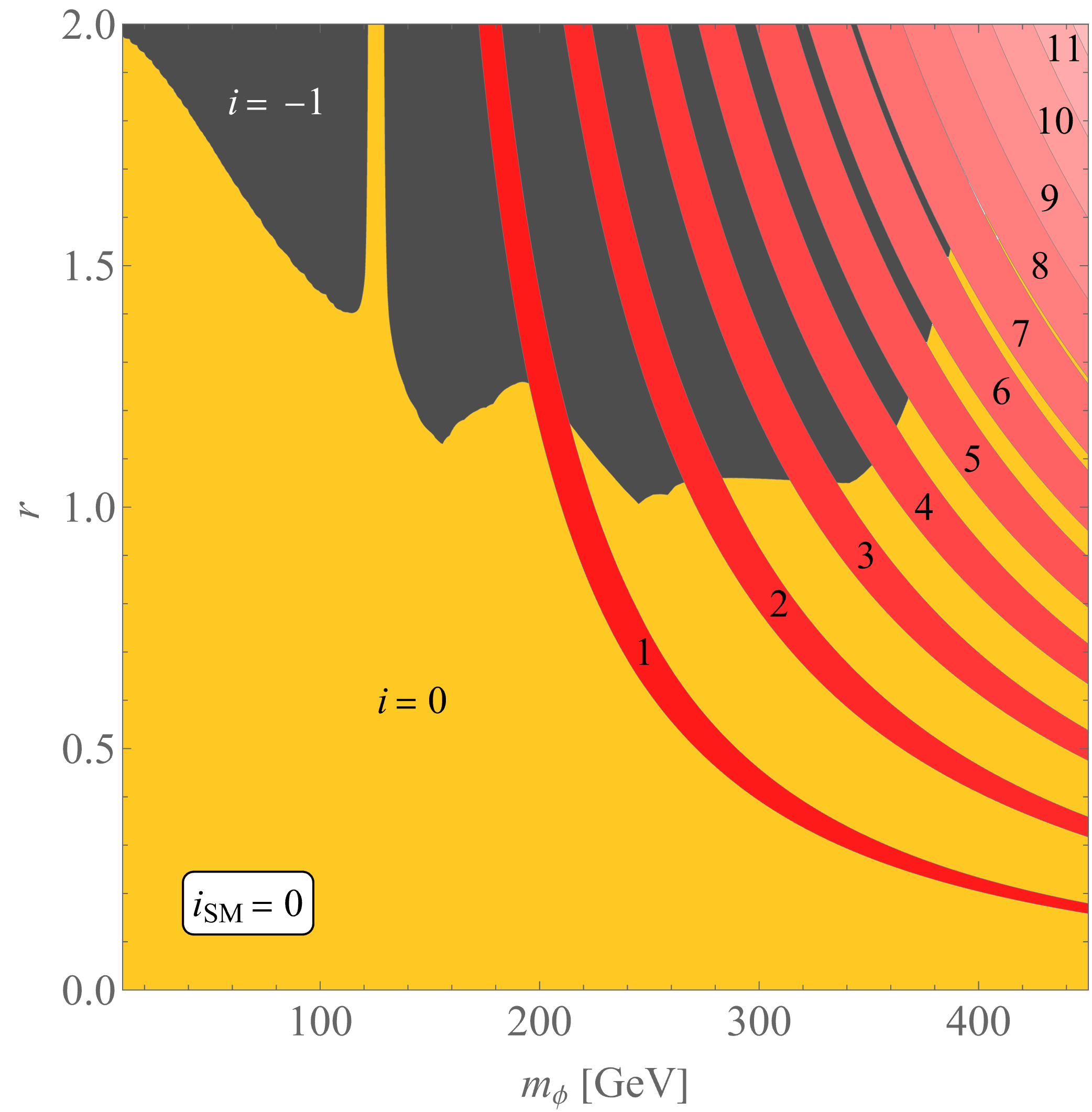} 
\end{center}
\caption{Parameter space for the original $N$-naturalness model ($i_{\rm SM} = 0)$. Colored regions denote the sectors that acquire the largest share of the energy density after reheating, which are labeled by sector index. 
For low reheaton masses, the SM $(i = 0)$ dominates for much of the parameter space (yellow). 
The lightest exotic sector $(i = -1)$  dominates at large values of $r$ and low $m_\phi$ where on-shell two-body reheaton decays to pairs of Higgs doublets are allowed (gray). 
As the reheaton mass increases, it will resonantly mix with the Higgs boson of one of heavier SM-like sectors ($i > 0)$, efficiently populating that sector (red bands). 
}
\label{fig:iSM-0}
\end{figure}

In the SM-like sectors electroweak symmetry breaking leads to $\phi-h_i$ mass mixing. The mixing angle is given by
\begin{equation}
\theta_i \simeq \frac{a v_i}{m_{h_i}^2 - m_\phi^2},
\label{eq:mixing}
\end{equation}
where $m_{h_i}$ is the physical Higgs mass for the sector. For a light reheaton, $m_\phi \ll m_{h_i}$, the mixing angle is approximately  
$\theta_i \approx a v_i/m^2_{h_i} \propto 1/m_{h_i}$.
Thus, the $\phi$ partial decay widths to SM-like sectors typically scale as $\Gamma_i \sim 1/m^2_{h_i}$, with decays to the lightest sector $i = 0$ being the largest. This is the basic feature which underlies the success of the original $N$-naturalness model. Fig.~\ref{fig:iSM-0} illustrates the $m_\phi - r$ parameter space for the $i_{\rm SM} = 0$ model, showing that across a wide range of light $m_\phi$ values, the SM obtains the largest fraction of the energy density (yellow region).

In the exotic sectors, electroweak symmetry breaking effects from QCD are negligible for reheaton decays. For all but the few lightest sectors, $m_\phi \ll m_{H_i}$, so reheaton decays proceed mainly via loops ($\phi \to W_i W_i,\, B_i B_i$) with widths $\Gamma_i \propto 1/m_{H_i}^4$, or through suppressed four-body decays ($\phi \to H_i^* H_i^*$). Consequently, heavier exotic sectors receive negligible energy density in the viable parameter space. Only the lightest exotic sectors may acquire a significant fraction of the reheaton’s energy density. As $r$ increases, the corresponding Higgs mass $m_{H_{i}}$ decreases, and when $m_{H_{i}} \lesssim m_\phi/2$, the reheaton can have a sizable branching ratio into this sector via two- or three-body decays ($\phi \to H_{i} H_{i}^{(*)}$). This feature is apparent in the large $r$, low $m_\phi$ portion of the parameter space in Fig.~\ref{fig:iSM-0} (gray region).  

Finally, it is important to note that in the SM-like sectors, the scaling $\Gamma_i \sim 1/m^2_{h_i}$ discussed above is strongly violated when the reheaton mass is close to the Higgs boson mass of a particular sector $i$, $m_\phi \sim m_{h_i}$. In such cases there is a resonant enhancement in $\phi-h_i$ mixing, as is evident from Eq.~(\ref{eq:mixing}), leading to the corresponding enhancement in the decays of the reheaton to the SM-like sector $i$. This effect explains why the reheaton dominantly decays into the heavier SM-like sectors as $m_\phi$ increases, as is clearly observed in  in Fig.~\ref{fig:iSM-0} (red bands). 

We stress that Fig.~\ref{fig:iSM-0} indicates only which sector acquires the dominant fraction of the energy density and should not be interpreted as the region consistent with observational constraints. Precision limits on additional relativistic species, quantified by $\Delta N_{\rm eff}$, exclude portions of the $i_{\rm SM} = 0$  parameter space (shown in yellow). The figure is intended to illustrate the overall structure of the parameter space and the sector favored by reheaton decays.

\section{$i$-incidental $N$-Naturalness}
\label{sec:ii-NN}

We now introduce $i$-incidental $N$-naturalness. The  construction is identical in all respects to the original $N$-naturalness model reviewed in the previous section, with the sole modification that sector $i_{\rm SM}$ (with $i_{\rm SM}> 0$) is to be identified with our SM. 
Using Eq.~(\ref{eq:mH-distribution-0}), the SM Higgs mass-squared parameter is written as 
\begin{equation}
m^2_{H_{i_{\rm SM}}}  \equiv  \mu^2 = - \frac{\Lambda_H^2}{N} (2 i_{\rm SM} + r ) \approx  - (88 \, {\rm GeV})^2.
\label{eq:mHSq-iNN}
\end{equation}
Then for the sector $i$ we can write
\begin{equation}
m_{H_i}^2 = \mu^2 \left(\frac{2 i +r}{2 i_{\rm SM} + r} \right), ~~~~~~- \frac{N}{2} \leq i \leq  \frac{N}{2}.
\label{eq:mH-distribution-1}
\end{equation}
Notably, the theories for different $i_{\rm SM}$ are not continuously connected. This is easily seen by the spectra of the theories. For a given $i_{\rm SM}$ there are $i_{\rm SM}$ SM-like Higgs particles lighter
than 125 GeV.

To address the hierarchy problem, we require a mechanism that cosmologically populates one of the SM-like sectors. In the original $N$-naturalness construction, where $i_{\rm SM} = 0$,  this occurs through the kinematic enhancement of reheaton decays into the lightest SM-like sector, as described in the previous section. What about the case $i_{\rm SM} >0$? 
The basic mechanism already presents itself in Fig.~\ref{fig:iSM-0} (red bands), where we observe that the scalar reheaton efficiently populates whichever SM-like sector $i > 0$ has a Higgs boson with mass closest to $m_\phi$ due to resonant $\phi - h_i$ mixing, see Eq.~(\ref{eq:mixing}). More generally, when  $m_\phi \gg v_{i = 0}$, the reheaton mass  necessarily  lies near the Higgs mass of {\it some} sector with $i > 0$ because the $m_{h_i}$ spectrum is dense (i.e., closely spaced). A small upward shift in  $m_\phi$ simply moves the reheaton off resonance with $h_i$ and on resonance with the Higgs boson from the next sector $h_{i+1}$. Thus, rather than being a coincidence, $m_{\phi} \approx  m_{h_i}$ for {\it some} $i$ is an unavoidable structural consequence of having many sectors.
Consequently, the reheaton dominantly decays into the sector whose Higgs mass is closest to its own mass. Provided the reheaton is light compared to the cutoff, with
\begin{equation}
v_{i = 0} \ll m_\phi \ll \Lambda_H,
\end{equation}
this resonantly populated sector will have an electroweak scale far below the cutoff, as required for a natural solution to the hierarchy problem. 

\begin{figure}[t]
\begin{center}
\includegraphics[width=0.475\textwidth]{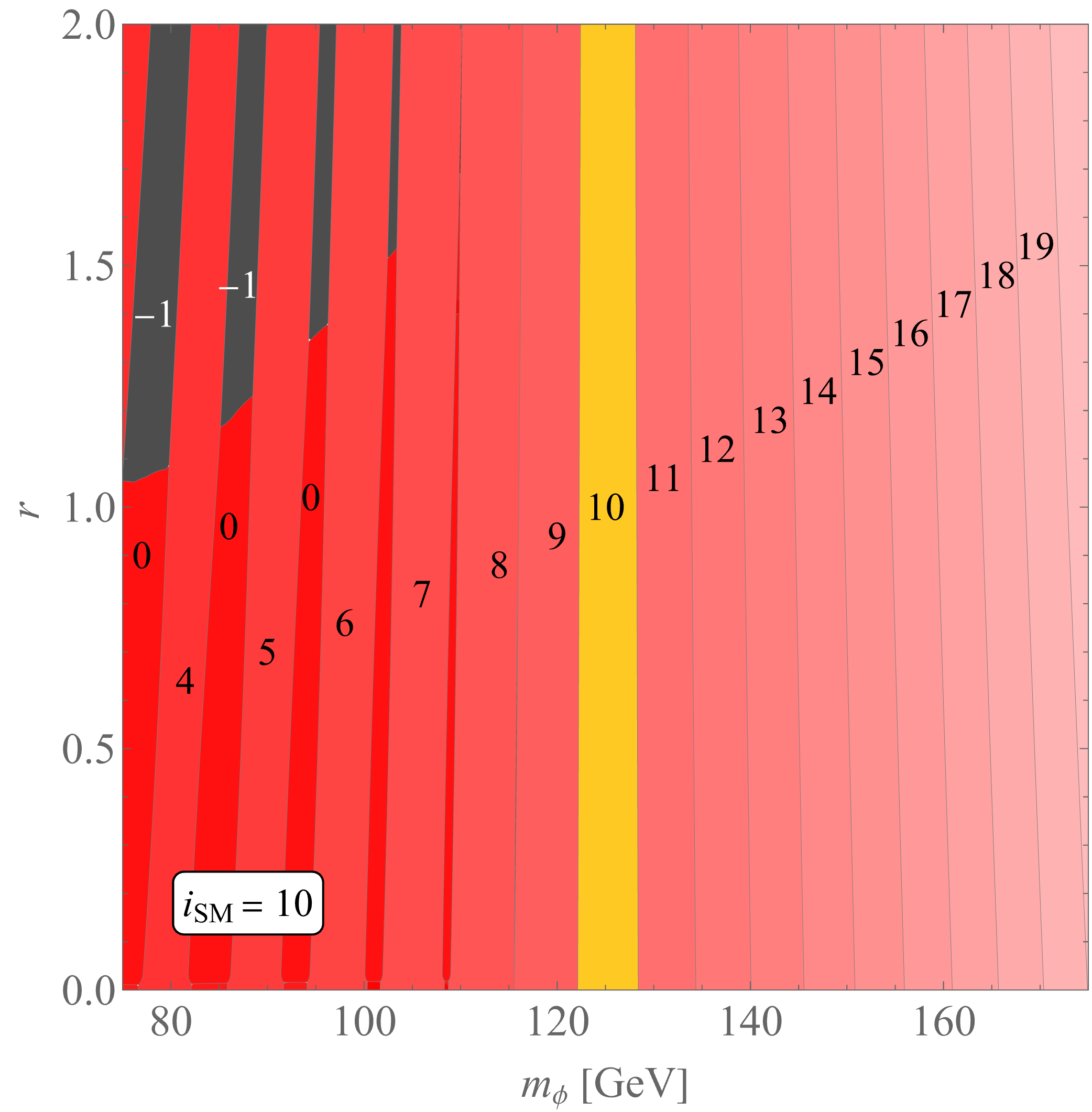} ~
\includegraphics[width=0.49\textwidth]{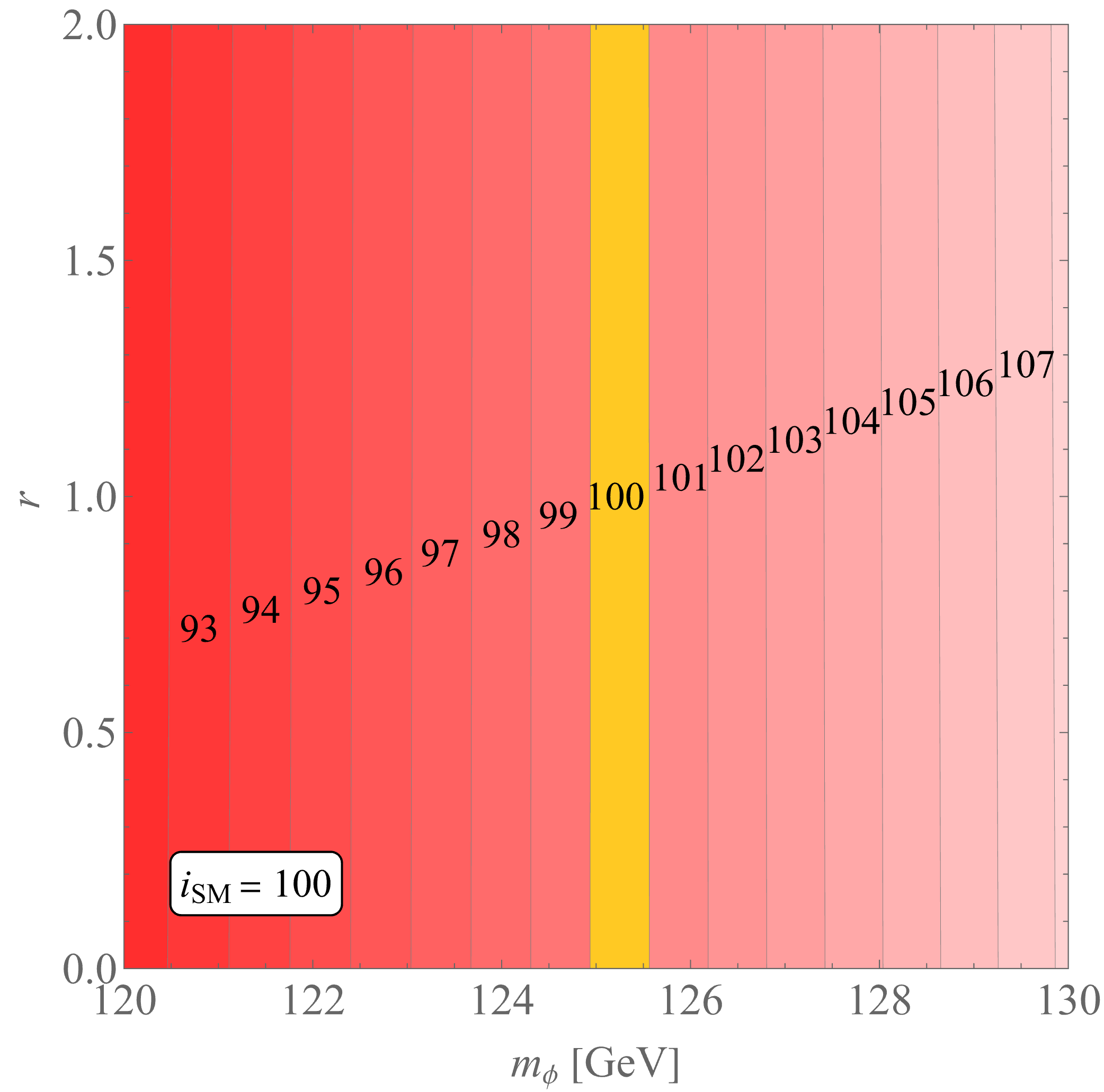} 
\end{center}
\caption{Parameter space for $i$-incidental $N$-naturalness for $i_{\rm SM} = 10$ (left panel) and $i_{\rm SM} = 100$ (right panel). 
As in Fig.~\ref{fig:iSM-0}, colored regions denote the sectors that acquire the largest share of the energy density after reheating, which are labeled by sector index. 
The reheaton predominantly populates our SM due to resonant mixing with the SM Higgs when $m_\phi \approx 125$ GeV (yellow). As the reheaton mass is varied away from the SM Higgs mass, it undergoes strong resonant mixing with the Higgs boson of another SM-like sector $i$ (red bands), thus dominantly populating that sector. 
Only for relatively light reheatons can it dominantly populate the lightest SM-like sector ($i = 0$) or the lightest exotic sector ($i = -1$) (gray). 
}
\label{fig:iSM-10-100}
\end{figure}

Therefore, as in the original $N$-naturalness model, this construction provides a dynamical mechanism that preferentially reheats one light sector over all others, singling it out as the SM, while all other sectors are subdominant.
Fig.~\ref{fig:iSM-10-100} illustrates the $m_\phi - r$ parameter space for the cases $i_{\rm SM} = 10$ (left) and $i_{\rm SM} = 100$  (right). As expected, the SM obtains the largest fraction of the energy density (yellow band) when the reheaton mass is in close proximity to the SM Higgs mass. Again, we emphasize that these figures merely show which sector receives the largest share of the reheaton energy density and are intended to illustrate the global structure of the parameter space. The nontrivial point, which we establish below in Sec.~\ref{sec:DNeff}, is that our sector receives the overwhelming majority of the reheaton energy density across a substantial portion of the SM ``resonance band'', ensuring compatibility with $\Delta N_{\rm eff}$ constraints for a wide range of $i_{\rm SM}$.

We require $\Lambda_H \gg v$ to solve the hierarchy problem. Inverting Eq.~(\ref{eq:mHSq-iNN}), 
\begin{equation}
\Lambda_H = |\mu| \left( \frac{N}{2 i_{\rm SM} + r}\right)^{1/2},
\end{equation}
we see that for fixed $N$, the cutoff decreases as $i_{\rm SM}$ increases and becomes less dependent on $r$. Equivalently, holding the cutoff fixed requires a larger number of sectors at higher $i_{\rm SM}$. 
For example, for $r\approx 1$, achieving $\Lambda_H \approx 10$ TeV  to address the little hierarchy problem requires approximately $10^4$, $10^5$, and $10^6$ sectors for $i_{\rm SM} = 0$, $10$, and $100$, respectively.

Given the constraint from the SM Higgs mass-squared parameter (\ref{eq:mHSq-iNN}), the minimal $N$-naturalness model is described by four parameters, which can be taken to be $\Lambda_H$, $r$, $m_\phi$, and $a$.
In $i$-incidental $N$-naturalness, $i_{\rm SM}$ is also to be considered as a free discrete parameter. 
We will typically consider the cutoff to be $\Lambda_H = 10$ TeV to address the little hierarchy problem, implying that the number of sectors $N$ increases as $i_{\rm SM}$ is increased. 
Interestingly, for large $i_{\rm SM}$ the phenomenological implications of the theory become rather insensitive to the parameter $r$.  This is already evident from Eq.~(\ref{eq:mHSq-iNN}) and from Fig.~\ref{fig:iSM-10-100}. 
This is in contrast to the original $N$-naturalness model where the physics is quite sensitive to $r$ (see Fig.~\ref{fig:iSM-0}) and small values of $r$ can be viewed as a tuning.
Furthermore, since the  Higgs portal coupling $a$ should be common to all the sectors, it will not impact the signatures of the model since it cancels out in the reheaton decay branching ratios. Besides $i_{\rm SM}$, the most important parameter is the reheaton mass $m_\phi$. Since the reheaton is nearly degenerate with the SM Higgs, it will often be useful to trade $m_\phi$ for the Higgs-reheaton mass splitting, 
\begin{equation}
\label{eq:h-phi-splitting}
\Delta m_{h\phi} \equiv m_h - m_\phi.
\end{equation}

\subsection{SM-like and Exotic Sectors}
\label{sec:sectors}

While the other sectors are indistinguishable from our SM in the UV, variations in their Higgs mass-squared parameters lead to pronounced differences at low energies.
Due to their negative Higgs squared masses, the SM-like sectors undergo electroweak symmetry breaking in the standard way, $\langle H_i \rangle \neq 0$, with the Higgs acquiring vacuum expectation values (VEVs) $v_{i}^2  = -m_{H_i}^2/\lambda = v^2 (2i+r)/(2 i_{\rm SM}+r)$, with $\lambda$ the common Higgs quartic coupling and $v= 246$ GeV the SM Higgs VEV. 

Instead, the exotic sectors have positive Higgs mass-squared parameters, so electroweak symmetry breaking is primarily driven by QCD, with the quark condensate  $\langle \overline q q \rangle_i \neq 0$ serving as the order parameter.
The exotic sector quarks obtain masses of order $m_{q_{i}} \sim y_q y_t \Lambda_{{\rm QCD}_i}^3/m_{H_i}^2$, thus they are all much lighter than corresponding confinement scale
$\Lambda_{{\rm QCD}_i} \sim {\cal O}(100 \, \rm MeV)$. 
This is particularly interesting since these exotic sectors, with six light quark flavors, may undergo first order phase chiral symmetry breaking phase transitions~\cite{Pisarski:1983ms}, leading to a corresponding stochastic gravitational wave signal. 
We will explore this further in Sec.~\ref{sec:grav-waves}. Similarly, the exotic sector spectrum features many light pions, light charged leptons, and very light neutrinos; see Ref.~\cite{Batell:2023wdb} for further details on the properties of the exotic sectors. 

Most of the SM-like (exotic) sectors have electroweak scales $v_i \gg v$ ($m_{H_i} \gg v$), but for $i_{\rm SM} > 0$, there will be order $i_{\rm SM}$ sectors that are light compared to us, in the sense that their Higgs degrees of freedom are light compared to our weak scale. Moreover, there will be some SM-like sectors that are nearby our sector in the sense that their Higgs boson masses $m_{h_i}$ are close to ours. As we will see below, these features impact the decays of the reheaton in important ways. 

\subsection{Reheating}
\label{sec:reheating}

Next, we turn to  the all-important reheating phase of $N$-naturalness. 
The reheaton is assumed to dominate the energy density after inflation and subsequently decays to all sectors via the Higgs portal coupling, Eq.~(\ref{eq:Higgs-portal}).
Each sector is reheated, receiving an energy density $\rho_i/\rho_{\rm SM} \simeq \Gamma_i/\Gamma_{\rm SM}$. The energy and entropy densities of each sector may be written as 
\begin{equation}
\rho_i = \frac{\pi^2}{30}\, g_{*\rho,i} \, \xi_i^4 \, T^4,
\quad\quad\quad
s_i = \frac{2\pi^2}{45} \, g_{*s,i} \, \xi_i^3 \, T^3,
\end{equation}
where $T$ is the SM temperature, $\xi_i \equiv T_i/T$ denotes the temperature of sector $i$ relative to that of the SM, and $g_{*\rho,i}$ ($g_{*s,i}$) represent  the effective number of relativistic (entropy) degrees of freedom in sector $i$. 
Sectors with lower energy densities will be colder since  $\rho_i \propto T_i^4$.\footnote{Throughout this work, subscripts on temperatures (and related quantities) label the sector, while superscripts indicate the cosmological epoch.} 
To avoid potential overproduction of stable relics, we assume that the baryon asymmetry in all other sectors is negligible~\cite{Arkani-Hamed:2016rle}. For related work on baryogenesis in $N$-naturalness, see Ref.~\cite{Easa:2022vcw}.

The reheating temperature $T^{\rm RH}$ of the SM plasma is set by the Higgs portal coupling $a$ and can therefore be treated as a free parameter. 
As in the original $N$-naturalness model, our reheating mechanism, relying on resonant $\phi-h$ mixing, requires electroweak symmetry to be broken in our sector during reheating, $T^{\rm RH} \lesssim v$. We will further assume that $T^{\rm RH} \lesssim {\cal O}(10 \,{\rm GeV})$ so that thermal corrections to the electroweak VEV, Higgs mass, and Higgs-reheaton mixing angle can be neglected. As we will discuss below, this implies that the Higgs-reheaton mixing angle must be smaller than about $10^{-7}$.
Furthermore, if $T^{\rm RH} \gtrsim \Lambda_{\rm QCD}$ then the exotic sectors will undergo chiral-symmetry breaking FOPTs, potentially providing an additional signature of  $N$-naturalness  in the form of a stochastic GW background (see Sec.~\ref{sec:grav-waves} below).

Assuming instantaneous reheating, the temperature ratio for the $i$th sector at reheating is given by
\begin{equation}
\label{eq:xi-RH}
\xi_i^{\rm RH} = \left[  \frac{  g^{\rm RH}_{*\rho,{\rm SM}} }{  g^{\rm RH}_{*\rho,i}  } \frac{\Gamma_i}{\Gamma_{\rm SM}}    \right]^{1/4}.
\end{equation}
Sector temperatures at later times can then be determined by standard arguments, such as conservation of entropy. 
Thus Eq.~\ref{eq:xi-RH} shows that the cosmology is set by the reheaton decay widths, $\Gamma_i$, which in turn depend on the $N$-naturalness model parameters. In fact, for large $i_{\rm SM}$ the main parameter is simply $m_\phi$ since its proximity to our Higgs mass controls the resonance enhancement.  

For a SM-like sector $i$, the reheaton decays through its mixing with the corresponding Higgs $h_i$ from that sector to all kinematically open final states $\{ f \}$,
\begin{equation}
\label{eq:decay-SML}
\Gamma_{\phi \to \{ f \} } = \theta_i^2 \; \Gamma_{h \to \{ f \} }(m_h \rightarrow m_\phi, v \rightarrow v_i),
\end{equation}
where $\theta_i$ is the mixing angle between $\phi$ and $h_i$ (\ref{eq:mixing}) and $\Gamma_{h \to \{ f \}}(m_h,v)$ denotes the corresponding SM Higgs partial decay width. For later convenience, we define the ratio
\begin{equation}
\label{eq:width-ratio}
    R_{\Gamma_{h_i}} \equiv
    \frac{\Gamma_{h,{\rm tot}}(m_h \rightarrow m_\phi, v \rightarrow v_i)}{\Gamma_{h}},
\end{equation}
where $\Gamma_{h} \equiv \Gamma_{h,{\rm tot}}(m_h,v) = 4.1$ MeV is the SM Higgs decay width. 
Furthermore, if the reheaton is heavy compared to $h_i$ it may decay to pairs of Higgs bosons, 
\begin{equation}
\Gamma_{\phi \to h_i h_i} = \frac{a^2}{32\pi m_\phi} \sqrt{1-\frac{4m_{h_i}^2}{m_\phi^2}}.
\end{equation}
For light exotic sectors, with $m_\phi > 2 m_{H_{i}}$, the reheaton can decay to pairs of on-shell Higgs doublets,
\begin{equation}
\Gamma_{\phi \to H_{i} H_{i}^\dagger} = \frac{a^2}{8\pi m_\phi} \sqrt{1-\frac{4m_{H_{i}}^2}{m_\phi^2}}.
\label{eq:Gamma-HH}
\end{equation}
On the other hand, as already discussed in Sec.~\ref{sec:original-NN}, reheaton decays to heavier exotic sectors involve loops or three- or four-body decays and are highly suppressed. Further details regarding reheaton decays may be found in Appendix~A of Ref.~\cite{Batell:2023wdb}.

Note that for large $i_{\rm SM}$, the spacing of the Higgs masses nearby $i_{\rm SM}$ is 
\begin{equation}
\label{eq:higgs-splitting}
\Delta m_h \equiv m_{h_{i_{\rm SM}}} - m_{h_{i_{\rm SM}-1}} \approx \frac{m_h}{2 \, i_{\rm SM}}.
\end{equation}
This spacing becomes smaller than the Higgs width, $\Delta m_h < \Gamma_h$, for 
\begin{equation}
\label{eq:small-splitting}
i_{\rm SM} \gtrsim  \frac{m_h}{2\, \Gamma_h} \approx 10^4.
\end{equation}
In this regime, we must account for the effect of the Higgs width in computing the reheaton decays. This can be accomplished by replacing the mixing angle in   Eq.~(\ref{eq:decay-SML}) with the effective mixing angle\footnote{A more accurate treatment would entail the computation of the full momentum dependent propagator of the mixed two scalar system, see, e.g., Refs.~\cite{Fuchs:2016swt,Boyanovsky:2017esz,Sakurai:2022cki,LoChiatto:2024guj,Kamada:2024ntk} for some recent studies. However, we do not expect this to significantly change our conclusions here.}
\begin{equation}
\label{eq-theta-eff}
\theta_i^2 \rightarrow \theta^2_{i,{\rm eff}} \simeq \frac{a^2 v_i^2 }{(m_\phi^2 - m_{h_i}^2)^{2} + m^2_{h_i} \Gamma^2_{h_i}}.
\end{equation}
In fact, in this regime, the reheating mechanism fails due to the relative broadness of the Higgs width compared to the Higgs mass spacing, implying that the reheaton will efficiently populate many nearby sectors. Thus, $\Delta N_{\rm eff}$ is unavoidably large in this regime, leading to an upper bound on $i_{\rm SM}$. This will be examined in detail below. 
 
Assuming instantaneous reheating and that the reheaton dominantly decays to our sector, 
we can relate the reheaton-Higgs mixing angle to the reheating temperature, 
\begin{equation}
\label{eq:theta-TRH}
\theta_{i_{\rm SM}} \approx \frac{T^{\rm RH}}{ \sqrt{M_{\rm Pl} \Gamma_h}} 
\simeq 10^{-7} \left(\frac{ T^{\rm RH}}{10\, \rm GeV}   \right).
\end{equation}
As discussed above, we will assume that $T^{\rm RH} \lesssim 10$ GeV, so that thermal effects can be neglected. This implies that the mixing angle is extremely small, $\theta_{i_{\rm SM}}  \lesssim 10^{-7}$. 
Even if this requirement can be relaxed somewhat, the resonant reheating mechanism relies on our sector being in the broken electroweak symmetry phase, 
$T^{\rm RH} \lesssim 100$ GeV, in which case the mixing angle is required to satisfy $\theta_{i_{\rm SM}}  \lesssim 10^{-6}$. 
Since the Higgs portal coupling can be freely varied without impacting the reheaton branching ratios, it is always possible to satisfy this condition. On the other hand, such small mixing angles makes searching for the reheaton at colliders practically hopeless. 
We comment further on this possibility in Sec.~\ref{sec:collider}.

Our estimates below of $\Delta N_{\rm eff}$, relic abundances of stable massive particles, and gravitational wave spectra require as input the number of effective relativistic degrees of freedom in each sector at various cosmological epochs. These factors are determined numerically according to the spectrum of the sector and its temperature at the epoch under consideration. The latter is mainly set by $T^{\rm RH}$ and $m_\phi$ once $i_{\rm SM}$ is fixed. 

\section{Signatures and Constraints }
\label{sec:constraints}

In this section we examine the key signatures and constraints relevant for $N$-naturalness with $i_{\rm SM} > 0$. We discuss cosmological probes such as measurements of additional relativistic species ($\Delta N_{\rm eff}$) and overclosure bounds from relic stable massive particles, as well as possible stochastic GW signals from QCD FOPTs in the light exotic sectors. We also comment on the prospects and inherent challenges of probing the 125 GeV reheaton directly at high energy colliders, which would provide a smoking-gun signature of the scenario. 

\subsection{$\Delta N_{\rm eff}$}
\label{sec:DNeff}

One of the key predictions of $N$-naturalness is dark radiation from the other sectors, which can be encoded in the parameter $\Delta N_{\rm eff}$. 
The most stringent constraints on the model come from bounds on $\Delta N_{\rm eff}$  during the epoch of recombination. The precise limits depend on the choice of datasets and exhibit some model dependence, such as whether the dark radiation is free-streaming or interacting.
One of the standard bounds on free-streaming dark radiation comes from Planck, in particular ${\rm Planck+Lensing+BAO}$~\cite{Planck:2015fie},  
$\Delta N_{\rm eff}^{\rm CMB}  \leq 0.3$ (here and below bounds are reported at approximate $95\%$ confidence level). 
More recently, the Atacama Cosmology Telescope (ACT) collaboration obtained a limit of
$\Delta N_{\rm eff}^{\rm CMB} \leq 0.2$ by combining its Data Release 6 (DR6) with Planck and BBN data~\cite{AtacamaCosmologyTelescope:2025nti}. 
Looking ahead, the Simons Observatory aims to achieve a sensitivity of $\Delta N_{\rm eff}^{\rm CMB} \leq 0.05$~\cite{SimonsObservatory:2018koc}, while future CMB and high-redshift surveys have the potential to reach 
$\Delta N_{\rm eff}^{\rm CMB} \leq 0.02$~\cite{Sailer:2021yzm,MacInnis:2023vif,CMB-S4:2016ple}.
Some caution should be taken in applying these limits. The exotic sectors behave more like an interacting fluid, leading to a somewhat weaker bound of 
$\Delta N_{\rm eff}^{\rm CMB} \leq 0.45$~\cite{Schoneberg:2021qvd}.
Finally, there is the longstanding discrepancy between Planck and SH0ES measurements of the Hubble constant. Incorporating the SH0ES data~\cite{Riess:2020fzl} further relaxes the bound on interacting radiation to $\Delta N_{\rm eff}^{\rm CMB} \leq 0.7$~\cite{Blinov:2020hmc}. 
While comparable limits can be placed on $\Delta N_{\rm eff}$ during the big bang nucleosynthesis (BBN) epoch, $N$-naturalness typically predicts $\Delta N_{\rm eff}^{\rm CMB} > \Delta N_{\rm eff}^{\rm BBN}$.  

We compute $\Delta N_{\rm eff}^{\rm CMB}$  near recombination at a SM temperature $T^{\rm CMB} = 0.3$ eV by summing the contributions from all other sectors,
\begin{align}
\label{eq:DNeff0}
\Delta N_{\rm eff}^{\rm CMB} =  \sum_{i \neq i_{\rm SM}} \Delta N_{{\rm eff},i}^{\rm CMB},
\end{align}
where the contribution from sector $i$ is 
\begin{align}
\label{eq:DNeff1}
\Delta N_{{\rm eff},i}^{\rm CMB} = \frac{8}{7} \left( \frac{11}{4} \right)^{4/3} \left[\frac{g_{*\rho,i}^{\rm CMB}}{2} \right](\xi^{\rm CMB}_i)^4.
\end{align}
It is straightforward to relate  
$\xi_i^{\rm CMB}$ 
to 
$\xi_i^{\rm RH}$ 
in Eq.~\eqref{eq:xi-RH}. The latter is determined by the reheaton decay width ratio $\Gamma_i/\Gamma_{\rm SM}$ and thus by the parameters $m_\phi$, $r$, and $i_{\rm SM}$. 
We will do this in several steps to allow for the possibility of entropy production due to a QCD FOPT in the $i$th sector. 
The first step is to connect  $\xi_{i}^{\rm CMB}$ to $\xi_{i}^{{\rm rh},i}$, corresponding to the reheating of sector $i$ following its FOPT, making use of the fact that the total entropy in both the SM sector and the SM-like sectors is conserved between the epochs. Second, we allow for the possibility of a change in the $i$th sector temperature during the phase transition between the time of percolation and reheating,
\begin{align}
\label{eq:Trhi}
T_{i}^{{\rm rh},i}
= T_{i}^{{\rm perc},i} \,
 (1+ \alpha_{i})^{1/4} \left[\frac{ g_{*\rho,i}^{{\rm perc},i}  }{ g_{*\rho,i}^{{\rm rh},i}  } \right]^{1/4} .
\end{align}
Here $\alpha_{i}$ characterizes the strength of the FOPT; see Sec.~\ref{sec:grav-waves} below for more details. This relation is obtained using energy conservation and assuming an instantaneous transition from percolation to reheating; see Ref.~\cite{Batell:2023wdb} for further discussion. 
Eq.~(\ref{eq:Trhi}) then provides  a relation between  $\xi_{i}^{{\rm rh},i}$ and $\xi_{i}^{{\rm perc},i}$. 
For the final step, we again rely on  entropy conservation to relate the $\xi_{i}^{{\rm perc},i}$ to $\xi_{i}^{\rm RH}$. 
The final result for $\Delta N_{{\rm eff},i}^{\rm CMB}$ is
\begin{align}
\label{eq:DNeff-i}
\Delta N_{{\rm eff},i}^{\rm CMB}  = \frac{8}{7} \left( \frac{11}{4} \right)^{4/3}
 \left[\frac{g_{*\rho,{\rm SM}}^{\rm RH}}{2} \right]  
 \left[ \frac{ g_{*s,{\rm SM}}^{\rm CMB} }{ g_{*s,{\rm SM}}^{\rm RH} }   \right]^{4/3} 
  \left[\frac{ g_{*\rho,i}^{\rm CMB}  }{ g_{*\rho,i}^{\rm RH}  } \right]  
   \left[\frac{ g_{*s,i}^{\rm RH}   }{ g_{*s,i}^{\rm CMB}  } \right]^{4/3} 
 D^{4/3}_{s,i}\,
 \frac{\Gamma_{i}}{\Gamma_{\rm SM}}.
\end{align}
where the possibility of entropy production from a QCD FOPT in the $i$th sector is encoded in the entropy density ratio, 
\begin{equation}
\label{eq-DSi}
D_{s,i} \equiv \frac{s_{i}^{{\rm rh},i}}{s_{i}^{{\rm perc},i}} = \frac{ g_{*s,i}^{{\rm rh},i} \,
(T_{i}^{{\rm rh},i})^3 }{ g_{*s,i}^{{\rm perc},i} \, (T_{i}^{{\rm perc},i})^3} 
=(1+ \alpha_{i})^{3/4} \left[\frac{ g_{*s,i}^{{\rm rh},i} }{ g_{*s,i}^{{\rm perc},i} }\right] \left[\frac{ g_{*\rho,i}^{{\rm perc},i}}{ g_{*\rho,i}^{{\rm rh},i} } \right]^{3/4},
\end{equation}
If the phase transitions are weak, then $D_{s,i} \approx 1$. Furthermore, the factors involving $g_{*}$  in Eq.~(\ref{eq:DNeff-i}) typically give only small corrections of order $10\%$. 

An approximate expression for $\Delta N_{\rm eff}$ can be obtained by decomposing it into three components: (i) $\Delta N_{\rm eff,Ex}$ from light exotic sectors, (ii) $\Delta N_{\rm eff,SML}$ from light SM-like sectors, and (iii) $\Delta N_{\rm eff,Res}$ from SM-like sectors $i$ in the vicinity of $i_{\rm SM}$.
The first two terms can be appreciable due to light Higgs modes and the absence of kinematic suppression in reheaton decays.
The third contribution is also important, since nearby SM-like sectors experience resonant $\phi-h_i$ mixing, albeit weaker than in our own sector. 
Using the formulae for the reheaton decay widths given earlier, the sums in each region can be approximated by integrals, allowing us to derive semi-analytic expressions for the different contributions to $\Delta N_{\rm eff}$; see Appendix~\ref{sec:sums} for additional technical details of this procedure. 
The contribution from light exotic sectors is 
\begin{equation}
\label{eq:DNeff-exotic}
\Delta N_{\rm eff,Ex} \approx  \frac{8}{7} \left( \frac{11}{4} \right)^{4/3}  \frac{m_h^3  }{12 \, \pi \, v^2 \, \Gamma_{h}}  \left(\frac{\Delta m_{h\phi}}{m_h}\right)^2  (2 i_{\rm SM}+r),
\end{equation}
where $\Delta m_{h\phi}$ is the Higgs-reheaton mass splitting, Eq.~(\ref{eq:h-phi-splitting}).
Due to the Goldstone boson equivalence theorem, the contribution from light SM-like sectors is approximately equal to the first contribution from the light exotic sectors,  
$\Delta N_{\rm eff,SML} \approx 1.33 \Delta N_{\rm eff,Ex}$,  where the ${\cal O}(1)$ numerical difference arises due to electroweak symmetry breaking effects in the SM-like sectors. 
The final contribution from SM-like sectors near $i_{\rm SM}$ undergoing resonant $\phi-h_i$ mixing is given by
\begin{equation}
\label{eq:DNeff-Res}
\Delta N_{\rm eff,Res} \approx  \frac{8}{7} \left( \frac{11}{4} \right)^{4/3} \, 2 c \, \left(\frac{\Delta m_{h\phi}}{m_h}\right)^2 \,  (2 i_{\rm SM}+r)^2,
\end{equation}
where $c \approx 1.7$ is obtained from a numerical fit.
The total contribution is then 
\begin{equation}
\Delta N_{\rm eff} \approx \Delta N_{\rm eff,Res}+\Delta N_{\rm eff,SML} + \Delta N_{\rm eff,Ex} \simeq \Delta N_{\rm eff,Res}+2.33\Delta N_{\rm eff,Ex}.
\label{eq:DNeff-approx}
\end{equation}
These approximations work well when the Higgs-reheaton mass splitting is smaller than the spacing between neighboring Higgs masses but still larger than the Higgs width, $\Gamma_h \ll \Delta m_{h\phi} \ll m_h/(2 i_{\rm SM})$. 
This is possible for $i_{\rm SM} \ll m_h/(2\Gamma_h) \approx 10^4$ (see Eqs.~(\ref{eq:higgs-splitting}),(\ref{eq:small-splitting})).

For larger values of $i_{\rm SM}$ the spacing of Higgs masses nearby our sector becomes smaller than the Higgs width (see Eq.~(\ref{eq:small-splitting})), which must then be accounted for in the reheaton decays using the effective mixing angle given in Eq.~(\ref{eq-theta-eff}).
In the regime where $\Delta m_h \ll \Gamma_h$ (or $i_{\rm SM}\gg 10^4$ from Eqs.~(\ref{eq:higgs-splitting},\ref{eq:small-splitting})), the  dominant contribution to $\Delta N_{\rm eff}$ comes from the resonance region, and an approximate expression for this contribution is 
\begin{equation}
\label{eq:DNeff-Res-1}
\Delta N_{\rm eff,Res} \approx \frac{8}{7} \left( \frac{11}{4} \right)^{4/3}\left( \frac{\pi \,\Gamma_h}{m_h}  \, i_{\rm SM} -1 \right).
\end{equation}
Requiring $\Delta N_{\rm eff} \lesssim 0.3$, Eq.~(\ref{eq:DNeff-Res-1}) indicates an upper bound on $i_{\rm SM}$ of order $10^4$. Numerically, the actual bound is found to be somewhat tighter, $i_{\rm SM} \lesssim 4 \times 10^3$. This is obtained by setting the $m_\phi = m_h$, which maximizes the reheaton-Higgs mixing angle in our sector. 

\begin{figure} [htb]
 \includegraphics[width=0.43\linewidth]{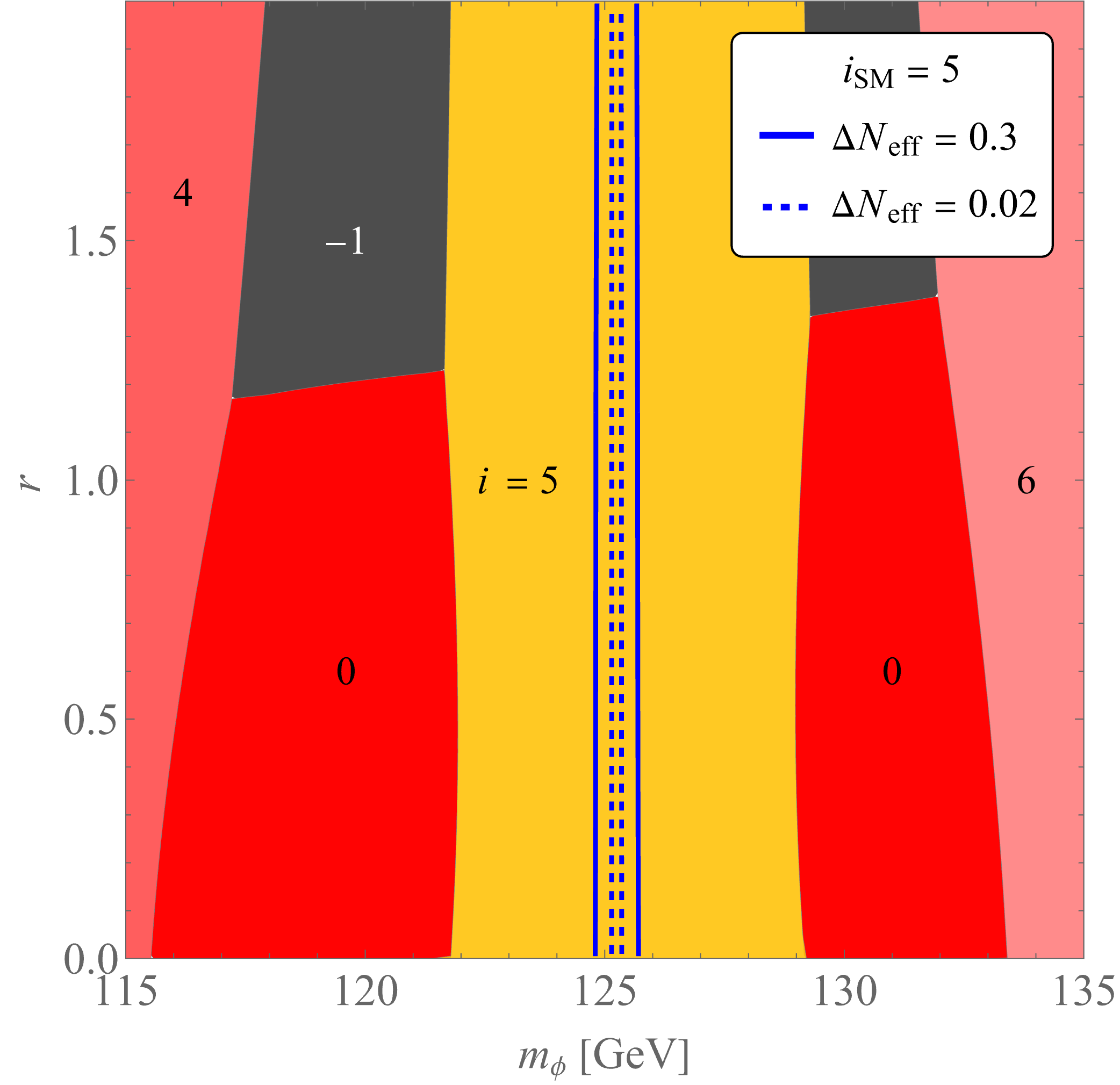} \quad\quad
 \includegraphics[width=0.43\linewidth]{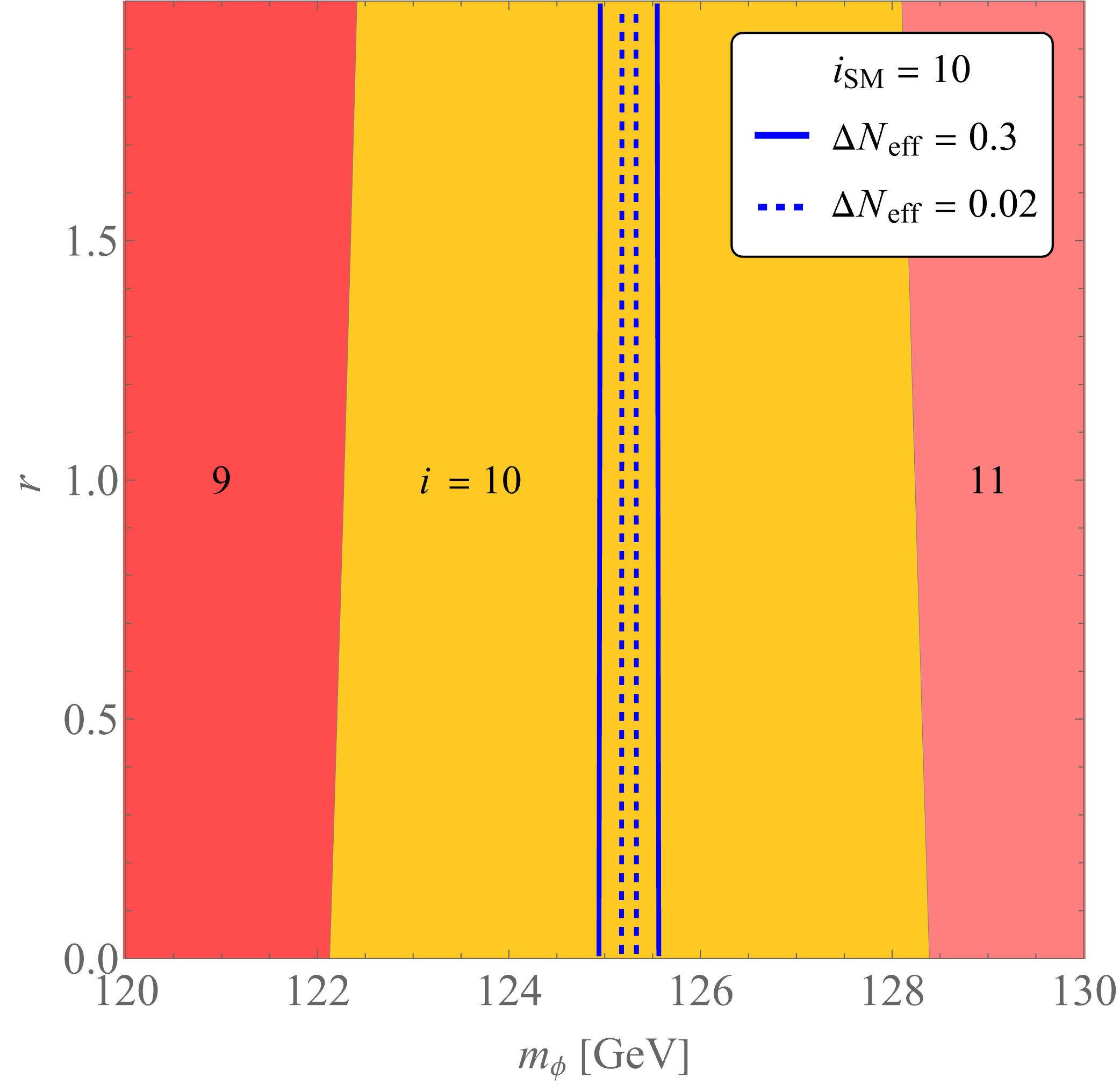} \\
 \includegraphics[width=0.43\linewidth]{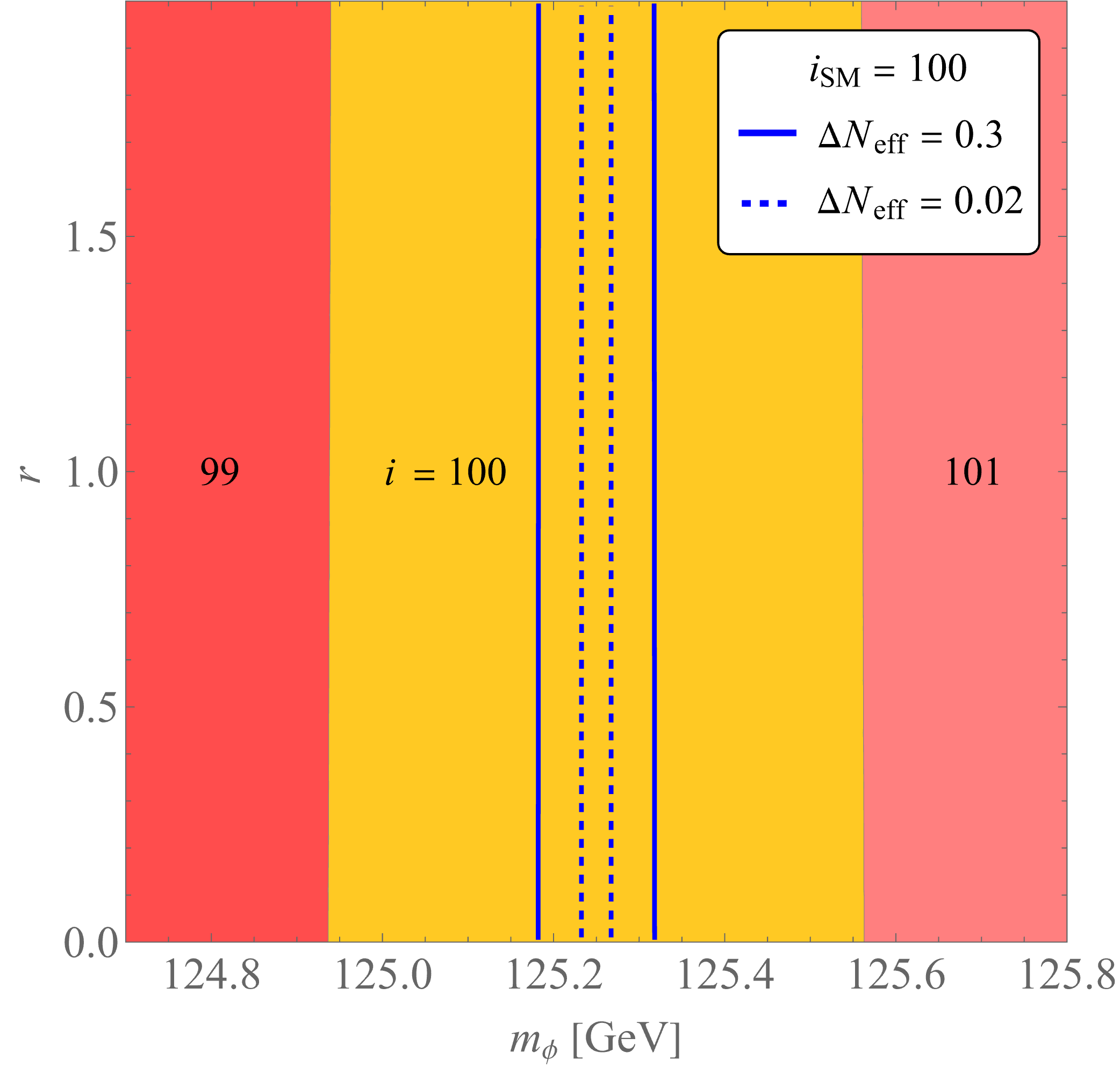}  \quad\quad
 \caption{Constraints on $N$-naturalness from $\Delta N_{\rm eff}^{\rm CMB}$ measurements for $i_{\rm SM} = 5$ (top left), $10$ (top right), $100$ (bottom). In each plot we display the current ${\rm Planck+Lensing+BAO}$ bound,  $\Delta N_{\rm eff}^{\rm CMB} < 0.3$ ~\cite{Planck:2015fie} (allowed region lies between the solid blue lines) and the projected sensitiviy from future CMB and large redshift surveys 
 $\Delta N_{\rm eff}^{\rm CMB} < 0.02$~\cite{Sailer:2021yzm,MacInnis:2023vif,CMB-S4:2016ple} (dashed blue lines). As in Figs.~\ref{fig:iSM-0},\ref{fig:iSM-10-100}, colored shaded regions denote the sectors that acquire the largest share of the energy density after reheating, which are labeled by sector index. }
 \label{fig:DNefff}
\end{figure}

In Fig.~\ref{fig:DNefff} we show the current bounds and future sensitivity to $\Delta N_{\rm eff}^{\rm CMB}$ measurements for three benchmarks: $i_{\rm SM} = 5$, $10$, and $100$. The allowed parameter region lies between the two solid blue lines, where the reheaton-Higgs resonant mixing is strong.  We see that a sizable portion of parameter space, especially for larger  $i_{\rm SM}$, remains consistent with current limits, and that future $\Delta N_{\rm eff}^{\rm CMB}$ measurements will probe much of the remaining viable region.

\begin{figure} [t]
 \includegraphics[width=0.6\linewidth]{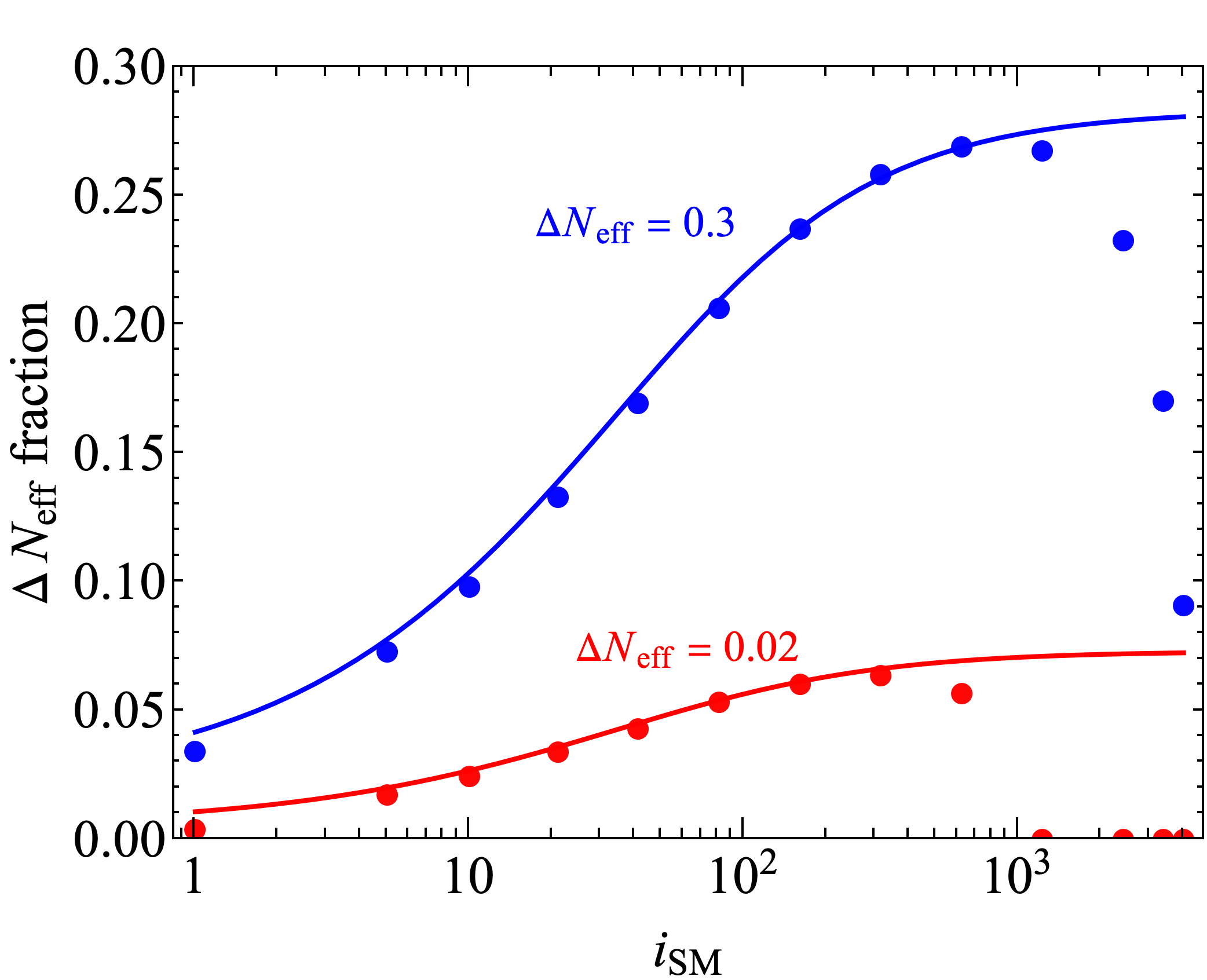}
 \caption{Fraction of the SM band that satisfies the bound $\Delta N_{\rm eff}^{\rm CMB}<0.3$  from ${\rm Planck+Lensing+BAO}$ (blue dots) and the sensitivity of future surveys $\Delta N_{\rm eff}^{\rm CMB}<0.02$ (red dots) 
 Also shown is the approximate semi-analytic estimate of the band valid for large $i_{\rm SM}$, Eq.~(\ref{eq:fraction}) (corresponding solid blue and red curves).}
 \label{fig:fraction}
\end{figure}

As these figures illustrate, the resonant reheating mechanism requires the reheaton mass to be sufficiently close to the SM Higgs mass. 
Let us define the SM resonance band to be the mass interval between the midpoints separating our Higgs from its nearest lighter and heavier neighbors,
\begin{equation}
\frac{1}{2} ( m_{h_{i_{\rm SM}}} + m_{ h_{ i_{ \rm SM}-1}} ) < m_\phi <  \frac{1}{2} (  m_{h_{i_{\rm SM}}} + m_{ h_{ i_{ \rm SM} + 1}} ).
\end{equation}
Note that the width of this band is approximately $m_h/(2 i_{\rm SM}+ r)$ for large $i_{\rm SM}$.
An interesting question is what fraction of the SM band respects the constraints imposed by $\Delta N_{\rm eff}^{\rm CMB}$ measurements. 
In Fig.~\ref{fig:fraction} we show this fraction as a function of $i_{\rm SM}$ (blue dots). We see that for small $i_{\rm SM}$, less than $\approx 10$, this fraction is relatively small, at the level of a few percent. 
The reason is that for small $i_{\rm SM}$ the width of the SM band is relatively broad, such that near the edges of the band the reheaton readily decays into the lightest SM-like sector 
$i = 0$ (for small $r$) or lightest exotic sector $i = -1$ (for large $r$). This feature is evident from Fig.~\ref{fig:DNefff} (top left) for the case $i_{\rm SM} = 5$.  
Instead, as $i_{\rm SM}$ increases beyond $\approx$ 10, the allowed fraction grows, reaching nearly one quarter for $i_{\rm SM}$ of a few hundred. 
In this regime the SM band shrinks, while resonant $\phi-h$ mixing remains strong across a significant portion of it. 
See Fig.~\ref{fig:DNefff} (bottom) for the example of $i_{\rm SM} = 100$.  
Finally, for very large $i_{\rm SM}$,  the Higgs mass spacing becomes comparable to the Higgs width, 
allowing the reheaton to mix resonantly with many nearby sectors and populate them efficiently. Consequently, the allowed fraction rapidly falls to zero once $i_{\rm SM}$ exceeds a few thousand.  

An approximate semi-analytic estimate of this fraction, valid for large $i_{\rm SM}$, is obtained using 
Eqs.~(\ref{eq:DNeff-approx},\ref{eq:DNeff-exotic},\ref{eq:DNeff-Res-1}) together with the SM band width $m_h/(2 i_{\rm SM}+ r)$,
\begin{equation}
\label{eq:fraction}
{\rm fraction}(i_{\rm SM},\Delta N_{\rm eff}) \approx \left\{ \frac{4 (2 i_{\rm SM}+r)  \Delta N_{\rm eff}  }{  \tfrac{8}{7} \left(\frac{11}{4} \right)^{4/3}  \left[ 2 c  (2 i_{\rm SM}+r)  + 2.33 m_h^3 /(12 \pi v^2 \Gamma_h)    \right]     }  \right\}^{1/2}.
\end{equation}
 This estimate is displayed in Fig.~\ref{fig:fraction} (red curve) and shows good agreement with the numerical estimate for $i_{\rm SM}$ less than ${\cal O}(1000)$. 

\subsection{Massive stable relics}
\label{sec:relics}

Here we assess how overclosure by massive, stable relics constrains the number of sectors $N$. As in Ref.~\cite{Arkani-Hamed:2016rle},  we find that
the dominant contribution comes from the freeze-out abundance of stable electrons and positrons residing in the SM-like sectors. 
We assume that the baryon asymmetry in other sectors is negligible and focus on the irreducible symmetric abundance.  
For a sector $i$ in which electrons have thermalized, 
$e^+ e^-$ annihilation to photons freezes out when the temperature drops significantly below the electron mass of the sector. The freeze-out temperature $T_i^{{\rm F},i}$ is given by
\begin{equation}
\label{eq:xfi}
x_i^{{\rm F},i} \equiv \frac{m_{e_i}}{T_i^{{\rm F},i}} = \log\left[ 0.19 \, g_e  \, (g_{*\rho,{\rm SM}}^{{\rm F},i})^{-1/2}\, m_{e_i} \,M_{\rm Pl}\, \langle \sigma v \rangle_i  \right] + 2 \log \xi_i^{{\rm F},i}+ \frac{1}{2} \log x_i^{{\rm F},i}.
\end{equation}
Here $g_e = 4$, the cross section for electron - positron annihilation is
\begin{equation}
 \langle \sigma v \rangle_i  \approx \frac{\pi \alpha^2}{m_{e_i}^2},
 \end{equation}
 the electron mass in sector $i \geq 0$ is 
\begin{equation}
m_{e_i}  = m_e \left(\frac{2i+ r}{2 i_{\rm SM}+r}\right)^{1/2},
\end{equation}
and $\xi_i^{{\rm F},i} = T_i^{{\rm F},i}/T^{{\rm F},i}$ is the temperature ratio at freezeout. Concerning the latter, $\xi_i$ is nearly constant in time with overall value set by the reheating phase.  
Thus, in the following, for simplicity we take $\xi^{{\rm F},i}_i \approx \xi^{\rm RH}_i$ as determined above in Sec.~\ref{sec:reheating}. 

Assuming the electrons in a sector thermalize, their relic abundance is given by 
\begin{equation}
\Omega_{e,i}^0 \simeq \frac{ 0.34 \, (g_{*s,{\rm SM}}^{0} / g_{*s,{\rm SM}}^{{\rm F},i}) (g_{*\rho,{\rm SM}}^{{\rm F},i})^{1/2} \, \xi_i^{{\rm F},i}  \, x_i^{{\rm F},i} \,(T^0)^3}{\rho_c^0\, M_{\rm Pl} \, \langle \sigma v \rangle_i }.
\end{equation}
The total relic abundance is found by summing over all such sectors,  
\begin{align}
\Omega_{e}^0 & =   \sum_{i=1}^{N_{\phi}}{}^{\! \prime}  \, \Omega_{e,i}^0   \simeq 0.11 \frac{m_e^2 \, (T^0)^3 }{\rho_c^0\, M_{\rm Pl} \, \alpha^2}  
  \sum_{i=1}^{N_{\phi}}{}^{\! \prime}  \,\,
  \frac{g_{*s,{\rm SM}}^{0}}{ g_{*s,{\rm SM}}^{{\rm F},i}} (g_{*\rho,{\rm SM}}^{{\rm F},i})^{1/2} \, \xi_i^{{\rm F},i}  \, x_i^{{\rm F},i}
 \left[ \frac{2 i + r }{2 i_{\rm SM} + r}  \right],
\end{align} 
where the prime indicates omission of the term $i = i_{\rm SM}$ in the sum and $N_{\rm \phi}$ denotes the number of sectors with thermal populations of electrons. Assuming the Higgs-reheaton spacing is smaller than Higgs width, the sum can be performed as follows: 
\begin{align}
\label{eq:sum-relic}
\sum_{i=1}^{N_{\phi}}{}^{\! \prime}  \,\,
\frac{g_{*s,{\rm SM}}^{0}}{ g_{*s,{\rm SM}}^{{\rm F},i}} (g_{*\rho,{\rm SM}}^{{\rm F},i})^{1/2} \, \xi_i^{{\rm F},i}  \, x_i^{{\rm F},i}
\left[ \frac{2 i + r }{2 i_{\rm SM} + r}  \right] &~ \approx ~  i_{\rm SM}^{-3/4}  \left[\frac{ |\Delta m_{h\phi}| }{m_h}\right]^{1/2}  \sum_{i=1}^{N_{\phi}}{}^{\! \prime}  \,\, \gamma(i)   i^{3/4}  \nonumber \\
&~ \approx ~ \frac{4}{7}  i_{\rm SM}^{-3/4}  \left[\frac{|\Delta m_{h\phi}| }{m_h}\right]^{1/2}  \gamma(N_{\phi})  \, N^{7/4}_{\phi},~~~~~~~
\end{align}   
where we have introduced the slowly varying function $\gamma(i)$ appearing in the sum,
 \begin{equation}
 \gamma(i) \equiv \sqrt{2} \left[\frac{g_{*s,{\rm SM}}^{0}}{ g_{*s,{\rm SM}}^{{\rm F},i}}\right] [g_{*\rho,{\rm SM}}^{{\rm F},i}]^{1/2}     \left[ \frac{g_{*\rho,{\rm SM}}^{\rm RH}}{ g_{*\rho,i}^{\rm RH}} \right]^{1/4}  R^{1/4}_{\Gamma_{h_i}}  \, \, x_i^{{\rm F},i} .
 \end{equation} 
 where $R_{\Gamma_{h_i}}$ is defined in Eq.~(\ref{eq:width-ratio}).
 We then obtain 
\begin{align}
\Omega_{e}^0 & \approx 0.063 \frac{m_e^2 \, (T^0)^3 }{\rho_c^0\, M_{\rm Pl} \, \alpha^2}    i_{\rm SM}^{-3/4}  \left[\frac{|\Delta m_{h\phi}|}{m_h}\right]^{1/2}  \gamma(N_{\phi})  \, N^{7/4}_{\phi} \\
&\approx 10^{-2} \left( \frac{i_{\rm SM}}{10}\right)^{-3/4} \left( \frac{|\Delta m_{h\phi}|}{3 \, {\rm GeV}}\right)^{1/2}  
\left( \frac{N_{\phi}}{10^7}\right)^{7/4}. \nonumber
\end{align} 
Constraints on self interactions of dark matter from observations of the bullet cluster limit strongly self interacting dark matter to be $\lesssim 23\%$ of all of dark matter \cite{Randall:2008ppe}. Stronger constraints can be obtained from dark acoustic oscillations imprinted on the CMB \cite{Cyr-Racine:2012tfp, Cyr-Racine:2013fsa,Cyr-Racine:2021oal,Bansal:2021dfh}. However the above bounds are for a single thermalized dark sector containing a dark proton and a dark electron undergoing recombination. In our case, each sector contains an extremely small fraction of the DM density and contains dark electrons and positrons that wont undergo recombination. Moreover, the acoustic oscillations of different sectors will not be in phase. Thus, the above constraints won't directly apply. We postpone a detailed study of the acoustic oscillations of the various sectors in $N$-naturalness for future work. We conservatively require the abundance to be smaller than one percent of the DM energy density, $\Omega_{e}^0 \lesssim 0.01 \, \Omega_{\rm DM}$.
 This leads to an overclosure bound on the number of thermalized sectors
 \begin{equation}
 \label{eq:overclosure}
 N_\phi \lesssim 4 \times 10^6 \left( \frac{i_{\rm SM}}{10}\right)^{3/7}  \left( \frac{3 \, {\rm GeV}}{|\Delta m_{h\phi}|}\right)^{2/7} .
 \end{equation}  

We should also check how many sectors have thermalized electrons. For this to occur, we require
\begin{equation}
T_i^{\rm RH} \gtrsim T_i^{{\rm F},i}.
\end{equation}
Using Eqs.~(\ref{eq:xi-RH},\ref{eq:width-ratio},\ref{eq:xfi}),
we find that sectors with $i < N_{\rm th}$ have thermalized electrons, where
\begin{align}
N_{\rm th} & \lesssim  2^{2/3}  \left[ \frac{g_{*\rho,{\rm SM}}^{\rm RH}}{ g_{*\rho,N}^{\rm RH}} \right]^{1/3}  i_{\rm SM} \, R^{1/3}_{\Gamma_{h_N}}  \, \left[\frac{|\Delta m_{h\phi}| }{m_h}\right]^{2/3}\, \left[ x_N^{{\rm F},N} \frac{T^{\rm RH}}{m_e}  \right]^{4/3} \\
& \approx 2.4 \times 10^7  \left( \frac{i_{\rm SM}}{10}\right)  
 \left( \frac{| \Delta m_{h\phi}|}{3 \, {\rm GeV}}\right)^{2/3}  \left( \frac{T^{\rm RH}}{100 \, {\rm GeV}}\right)^{4/3}. \nonumber
 \end{align}
Since $N_{\rm th} > N_\phi$ for these parameters, the bound in Eq.~(\ref{eq:overclosure}) holds. 

The most important conclusion of this analysis is that a cutoff of $\Lambda_H \sim 10$ TeV to address the little hierarchy problem is comfortably compatible with the overclosure bound for any choice of $i_{\rm SM} \lesssim 10^3$ consistent with the bound on $\Delta N_{\rm eff}$ discussed earlier. 

Before moving on, we note that the symmetric baryon relic abundance is generally smaller than the electron abundance. The main reason is that the baryon abundance of sector $i$ grows more slowly with $i$ than the corresponding electron abundance. As seen from Eq.~(\ref{eq:sum-relic}), the electron abundance for sector $i$ is proportional to $i^{3/4}$, a consequence of  $\langle \sigma v\rangle_i^{-1} \sim m^2_{e_i} \sim i$ and $\xi_i \sim i^{-1/4}$. Instead, taking the baryon annihilation cross section to be  $\langle \sigma v\rangle_i^{-1} \sim m_{\pi_i}^{2} \sim \sqrt{i}$, we see that the baryon abundance for sector $i$ scales as $i^{1/4}$. In this discussion we have neglected the mild growth in $\Lambda_{{\rm QCD}_i}$ as $i$ increases, but this does not change the general conclusion that the baryon abundance from the other sectors is subdominant.

\subsection{Gravitational waves}
\label{sec:grav-waves}

Next, we examine the possibility that $N$-naturalness gives rise to a stochastic GW signal generated by first-order QCD chiral symmetry-breaking phase transitions in the other sectors.
While the QCD phase transition in the SM is understood to be a smooth crossover~\cite{Schmidt:2025ppy}, it has been argued that this phase transition is first order in QCD-like theories with three or more light flavors~\cite{Pisarski:1983ms}.\footnote{
The question of the order of the phase transition has been studied at various points over the past few decades using phenomenological models and lattice methods but remains unsettled; see for example Refs.~\cite{Iwasaki:1995ij,Butti:2003nu,Karsch:2003jg,Cuteri:2021ikv,Aarts:2023vsf}. 
For our purposes, we will adopt the standard assumption, motivated by Ref.~\cite{Pisarski:1983ms}, that exotic sectors with six light quarks undergo  first-order chiral phase transitions.} 
This condition is satisfied in the exotic sectors, where all six quarks are light compared to the confinement scale, and may also apply to the lightest SM-like sectors.

Ref.~\cite{Batell:2023wdb} investigated the GW signal in the original $N$-naturalness model, finding that, depending on the details of the phase transition dynamics, upcoming GW observatories may offer an additional probe of the model (see also Ref.~\cite{Archer-Smith:2019gzq}). 
In that model $(i_{\rm SM} = 0)$, the GW signal arises solely from the first exotic sector, resulting in the standard spectrum associated with a single first-order phase transition.
Instead, for the case of $i_{\rm SM} > 0$, there are approximately $i_{\rm SM}/2$ light exotic sectors that share a significant portion of the reheaton energy density, leading to the possibility of many exotic sector QCD phase transitions. 
This opens up the possibility, at least in principle, of a rather exotic GW spectrum built from the overlay of a large number of individual phase transition signals.

In the sectors exhibiting QCD FOPTs, the phase transition begins at the critical temperature  $T_{i}^{{\rm crit},i}$, defined by the point at which the true and false vacua are degenerate. 
We assume that the critical temperature is proportional to the confinement scale, thus $T_{i}^{{\rm crit},i}  = T^{\rm crit} (\Lambda_{{\rm QCD}_i} / \Lambda_{\rm QCD})$, where $T^{\rm crit}\approx 160$ MeV is the SM critical temperature~\cite{HotQCD:2018pds,Borsanyi:2020fev}.
Using one-loop running, the confinement scale in sector $i$ can be estimated as
\footnote{The estimate of $\Lambda_{{\rm QCD}_i}$ is modified slightly for SM-like sectors in which the the hierarchy between quark masses and the confinement scale differs from that of the SM; see, e.g., Ref.~\cite{Archer-Smith:2019gzq}.}
\begin{equation}
\label{eq:lambdaQCDi}
    \Lambda_{{\rm QCD}_i} \simeq 
    \begin{cases}
    (m_{t} \, m_{b} \, m_{c} )^{-2/21}(\Lambda_{\rm QCD})^{9/7}
    & {\rm for} ~~~ i < 0, \\
   \left(\displaystyle{\frac{m_{t_i}\, m_{b_i} \, m_{c_i} }{m_{t} \, m_{b} \, m_{c} }}\right)^{2/27}\Lambda_{{\rm QCD}} = \left(\displaystyle{\frac{2i+r}{2i_{\rm SM}+r}}\right)^{1/9}\Lambda_{\rm QCD}
    & {\rm for} ~~~  i \geq 0. \\
    \end{cases}
\end{equation}
Note that for the exotic sectors, $\Lambda_{{\rm QCD}_{i<0}}/\Lambda_{\rm QCD} \approx 0.3$, which is independent of $i$. 

Starting from the symmetric phase, bubbles of true vacuum nucleate and then expand as a result of the negative pressure from the potential difference between the true and false vacua. 
These bubbles eventually coalesce such that the $i$th sector ultimately transitions fully into the broken phase.  The nucleation temperature, $T_{i}^{{\rm nuc},i}$, denotes the point at which the first bubbles of the true vacuum appear. 
Once nucleated, bubbles expand until percolation is achieved, defined as the stage at which approximately $34\%$ of the volume has converted to the true vacuum, corresponding to temperature  $T_{i}^{\rm{perc},i} \lesssim T_{i}^{{\rm crit},i}$. 
The temperature of the SM plasma at percolation is then $T^{{\rm perc},i} = T_{i}^{{\rm perc},i}/\xi_{i}^{{\rm perc},i} $ where $\xi_{i}^{{\rm perc},i}$ is the temperature ratio between the $i$th sector and the SM just before percolation.

The strength of the $i$th sector phase transition is encoded in the parameters $\alpha_{i}$ and $\alpha_{{\rm  tot},i}$, defined as
\begin{align}
    \label{eq:alphas}
    \alpha_{i} = \frac{\Delta \theta_{i}}{\rho_{i}^{{\rm perc},i}} , ~~~~~    \alpha_{{\rm  tot},i} = \frac{\Delta \theta_{i}}{\rho_{\rm tot}^{{\rm perc},i}}  =   \alpha_{i} \frac{\rho_{i}^{{\rm perc},i}}{\rho_{\rm tot}^{{\rm perc},i}},
\end{align}
where $\Delta \theta_{i}$ denotes the change in the trace of the energy–momentum tensor across the transition, i.e., between the unbroken and broken phases. 
The characteristic inverse time scale of the phase transition in sector $i$ is given by the parameter $\beta_i$, which is related to the three-dimensional Euclidean bounce action $S_{3,i}$ via
\begin{equation}
\label{eq:betaoH}
\left(\frac{\beta}{H}\right)_i  = T_{i} \frac{d}{dT_{i}}  \frac{S_{3,i}}{T_{i}}   \bigg\vert_{T_{i}^{{\rm nuc},i}}.
\end{equation}

A stochastic GW signal is produced during a cosmological FOPT from several sources, including bubble wall collisions~\cite{Kosowsky:1992rz,Kosowsky:1992vn,Caprini:2007xq}, sounds waves in the plasma~\cite{Hindmarsh:2013xza,Hindmarsh:2015qta,Hindmarsh:2017gnf}, and magnetohydrodynamic turbulence~\cite{Kosowsky:2001xp,Dolgov:2002ra,Caprini:2009yp}. 
We will not consider the turbulence-induced GW component in this work as this source remains subject to substantial theoretical uncertainties~\cite{Breitbach:2018ddu,Huber:2008hg,Hindmarsh:2015qta,Caprini:2009yp}.
The relevant observable is the differential GW density parameter
\begin{equation}
\label{eq:GW-total}
\Omega_{{\rm GW}}(f) = \sum_i \Omega_{{\rm GW},i}(f),
\end{equation}
where  $\Omega_{{\rm GW},i}(f) = ({1}/{\rho_c}) \,d\rho_{{\rm GW},i}/{d \log f}$ denotes the contribution from sector $i$, with $f$ the GW frequency and $\rho_c$ the critical density. The emission spectra for the bubble wall collisions and sound waves sources can be parameterized as~\cite{Breitbach:2018ddu,Huber:2008hg,Hindmarsh:2015qta,Caprini:2009yp}
\begin{align}
\label{eq:OmegaGW-emission-BW}
\Omega^{\rm em}_{{\rm GW},{\rm BW},i}(f_{\rm em}) & =   \left[\frac{0.11 v^3_{{\rm w},i}}{0.42+v^3_{{\rm w},i}}\right]
 \,\left( \frac{\kappa_{{\rm BW},i}(\alpha_{i}) \, \alpha_{{\rm tot},i} }{1+ \alpha_{{\rm tot},i} }  \right)^2   
 \left( \frac{H }{\beta }  \right)^2_i  s_{\rm BW}(f_{\rm em}/f_{{\rm p},{\rm BW},i}), \\
\label{eq:OmegaGW-emission-SW}
\Omega^{\rm em}_{{\rm GW},{\rm SW},i}(f_{\rm em}) & = 0.159 \, v_{{\rm w},i} \,\left( \frac{\kappa_{{\rm SW},i}(\alpha_{i}) \, \alpha_{{\rm tot},i} }{1+ \alpha_{{\rm tot},i} }  \right)^2   \left( \frac{H }{\beta }  \right)_i  s_{\rm SW}(f_{\rm em}/f_{{\rm p},{\rm SW},i}),
\end{align}
where $f_{\rm em}$ is the frequency at emission. 
The main parameters determining the GW spectrum are the phase transition strength parameters $\alpha_{i}$ and $\alpha_{{\rm tot},i}$, the duration parameter $(\beta/H)_i$, and the wall velocity $v_{{\rm w},i}$. 
Furthermore, the efficiency factors $\kappa_{{\rm BW},i}$ and $\kappa_{{\rm SW},i}$ quantify
the fractions of released vacuum energy that go into bubble wall kinetic energy or into bulk fluid motion, respectively, for each source.
Our assumptions regarding these quantities
will be discussed below.  
The spectral shape functions in Eqs.~(\ref{eq:OmegaGW-emission-BW}) and (\ref{eq:OmegaGW-emission-SW}) 
are given by 
$s_{{\rm BW},i}(x) = 3.8 x^{2.8}/(1+2.8 x^{3.8})$ 
and 
$s_{{\rm SW},i}(x) = [7x^{3}/(4+3 \, x^{2})]^{7/2}$, respectively, 
while the associated peak frequencies are $f_{{\rm p},{\rm BW}, i} = 0.23 \, \beta_i$ and  $f_{{\rm p},{\rm SW}, i} = 0.53 \, \beta_i/v_{{\rm w},i}$.
We account for an additional suppression factor for the sound wave source, relevant at large $\beta/H$, which is given by $\Upsilon_{i} \simeq {\rm min} [1,3.38 \, {\rm max}[v_{{\rm w},i}, c_{\rm s}] (\beta / H)^{-1}_i \sqrt{(1+\alpha_{{\rm tot},i})/(\kappa_i(\alpha_{i}) \alpha_{{\rm tot},i}) }]$,
with $c_{\rm s} = 1/\sqrt{3}$ the sound speed in the relativistic plasma~\cite{Ellis:2020awk,Guo:2020grp}.

We identify the emission time with the percolation epoch, when a sizable fraction of the universe is occupied by bubbles of the true vacuum. To determine the present-day spectrum, it is necessary to include the effects of the cosmic expansion between emission and today, which redshifts both the GW energy density and frequency:
\begin{equation}
\label{eq:OmegaGW-today}
h^2 \, \Omega_{{\rm GW},i}^0(f) = h^2  {\cal R}_i \, \Omega^{\rm em}_{{\rm GW},i}\left( \frac{a^0}{a^{{\rm perc},i}} f \right).
\end{equation}
Here $\Omega_{{\rm GW},i}^0$ ($\Omega^{\rm em}_{{\rm GW},i}$) represents the GW spectrum today (at emission), $f$ is the frequency today, $a^0$ ($a^{{\rm perc},i}$) is the scale factor today (at percolation), and ${\cal R}_i$ is a redshift factor. The latter factors are given by  
\begin{align}
\frac{a^0}{a^{{\rm perc},i}} 
 & =  \left[ \frac{g_{*s,{\rm tot}}^{{\rm perc},i}}{g_{*s,{\rm tot}}^{\rm 0}}   \right]^{1/3} \frac{T^{{\rm perc},i}}{T^{0}}, \nonumber \\
 h^2 {\cal R}_i & = h^2 \left(\frac{a^{{\rm perc},i}}{a^0}\right)^4 \left(\frac{H^{{\rm perc},i}}{H^0}\right)^2   = h^2 \Omega_{\gamma}^0 \left[ \frac{g_{*\rho,{\rm tot}}^{{\rm perc},i}}{2}   \right]
 \left[ \frac{g_{*s,{\rm tot}}^{0}}{g_{*s,{\rm tot}}^{{\rm perc},i}}   \right]^{4/3},
\end{align}
where $T^0 = 2.725\, {\rm K}  \approx  0.235 \,{\rm meV}$ is the CMB temperature today, $H^{0}$ ($H^{{\rm perc},i}$) is the Hubble rate today (at percolation) with $H^0  = 100\, h \, {\rm km}\, {\rm Mpc}^{-1} \,{\rm s}^{-1}$ and 
$h^2\Omega_{\gamma}^0 \approx 2.47 \times 10^{-5}$ is the current photon density parameter. 
The various factors summing relativistic degrees of freedom are defined as
\begin{align}
    \label{eq:R}
    g_{*\rho,\rm tot}^{{\rm perc},i} & \simeq g_{*\rho, \rm SM}^{{\rm perc},i} + \sum_{i \neq i_{\rm SM}} g_{*\rho,i}^{{\rm perc},i}  (\xi_{i}^{{\rm perc},i})^{4}, \nonumber \\
    g_{*s,\rm tot}^{{\rm perc},i}  & \simeq g_{*s, \rm SM}^{{\rm perc},i}  + \sum_{i \neq i_{\rm SM}} g_{*s,i}^{{\rm perc},i}  (\xi_{i}^{{\rm perc},i} )^{3},   \nonumber \\
    g_{*s,\rm tot}^{0} & \simeq g_{*s, \rm SM}^{0}  + \sum_{i \neq i_{\rm SM}} g_{*s,i}^{0} (\xi_{i}^{0})^{3}.
\end{align}

The GW spectrum is governed primarily by the quantities $\alpha_{i}$, $(\beta/H)_i$, $v_{{\rm w},i}$, and the efficiency factors. 
In principle, all of these quantities could be extracted from the temperature-dependent effective potential characterizing the phase transition. 
However, in our setup the dynamics are strongly coupled, and a first-principles determination is currently not possible. 
While lattice methods would be the ideal tool for analyzing such transitions (see, e.g., Ref.~\cite{Aarts:2023vsf} for discussion), no existing simulations correspond directly to our scenario. 
Various phenomenological models have been employed in the literature to study FOPTs in QCD-like theories and estimate the resulting GW spectra~Refs.~\cite{Bai:2018dxf,Helmboldt:2019pan,Bigazzi:2020avc,Halverson:2020xpg,Huang:2020crf,Reichert:2021cvs}. 
While these analyses often point to relatively weak transitions with correspondingly long durations, significant uncertainties remain in the applicability of these phenomenological models.
As our primary goal is to illustrate the range of potential GW signals in $N$-naturalness, 
we refrain from modeling the effective potential explicitly and instead will investigate two representative benchmark scenarios, one with a runaway phase transition and another with a non-runaway transition:
\begin{itemize}
\setlength\itemsep{-1em}
\item Runaway scenario:
\vspace{-5pt}
\begin{align}
\alpha_{i} = 5,  ~~~~~ (\beta/H)_i = 10,  ~~~~~ v_{{\rm w},i} = 1, ~~~~~ \kappa_{{\rm BW},i} = 1,  ~~~~~ \kappa_{{\rm SW},i} = 0, 
\label{eq:runaway}
\end{align}
\item Non-runaway scenario:
\vspace{-5pt}
\begin{align}
\alpha_{i} = 0.3, ~~~(\beta/H)_i = 300, ~~~ v_{{\rm w},i} = \frac{1}{\sqrt{3}}, ~~~ \kappa_{{\rm BW},i} = 0, ~~~ \kappa_{{\rm SW},i} = \frac{\alpha^{2/5}_{i}}{0.017 +(0.997+\alpha_{i})^{2/5}}.
\label{eq:nonrunaway}
\end{align}
\end{itemize}   
For the non-runaway scenario, we utilize the numerical fitting function for the efficiency factor $\kappa$ from Ref.~\cite{Espinosa:2010hh}.
We note that our non-runaway scenario, Eq.~(\ref{eq:nonrunaway}), is broadly consistent with the findings from phenomenological studies of QCD-like phase transitions mentioned above. 
 
It is worth noting that Eqs.~(\ref{eq:runaway},\ref{eq:nonrunaway}) treat the phase transition parameters as universal across sectors.
For the exotic sectors, where all six quarks are much lighter than the confinement scale, this should be a good approximation.
Even so, their GW spectra differ because each sector receives a different energy density from the reheaton decays, affecting both the peak frequency and the amplitude.

\begin{figure}[htbp!]
\centering
\includegraphics[width=0.48\textwidth]{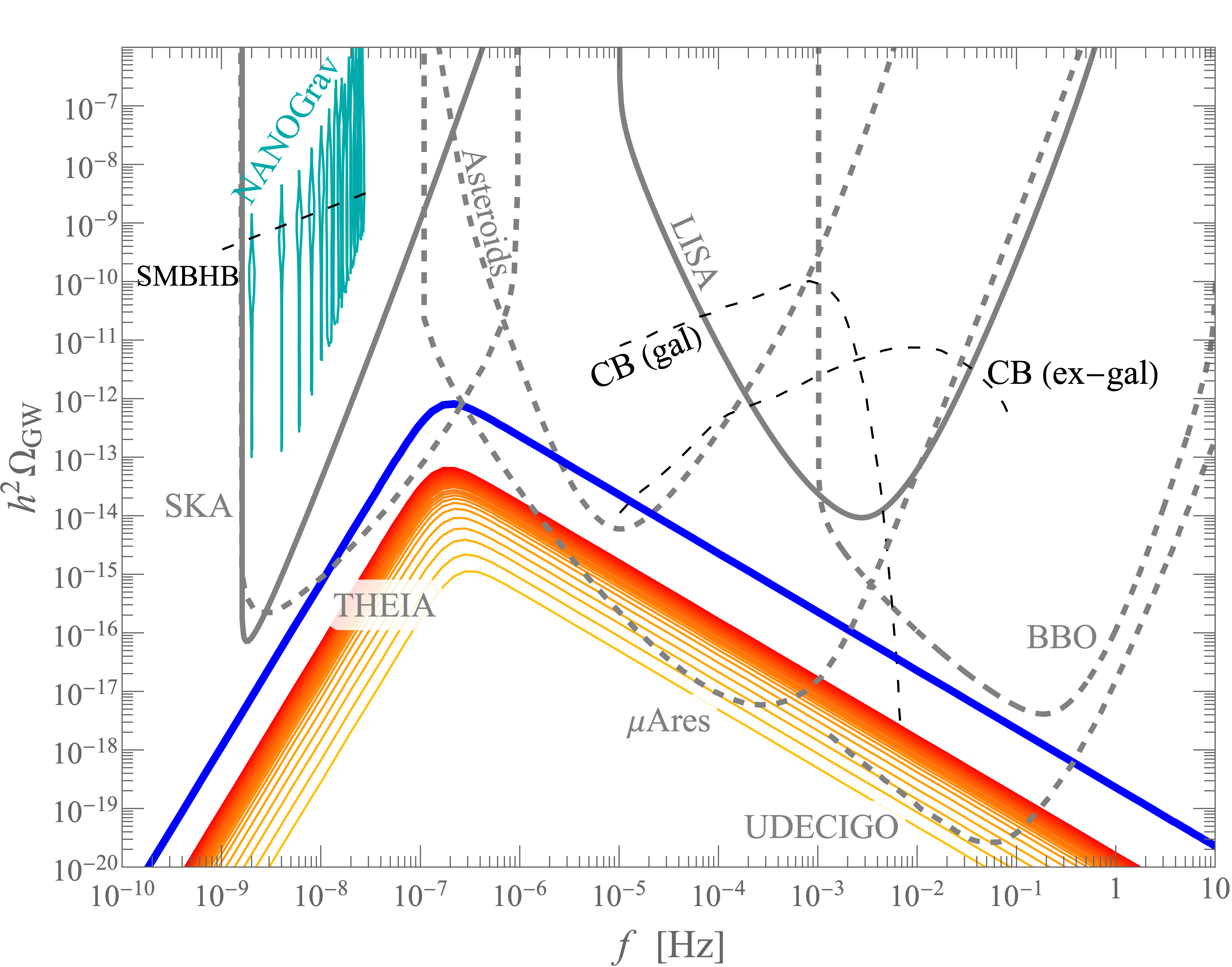}~
\includegraphics[width=0.48\textwidth]{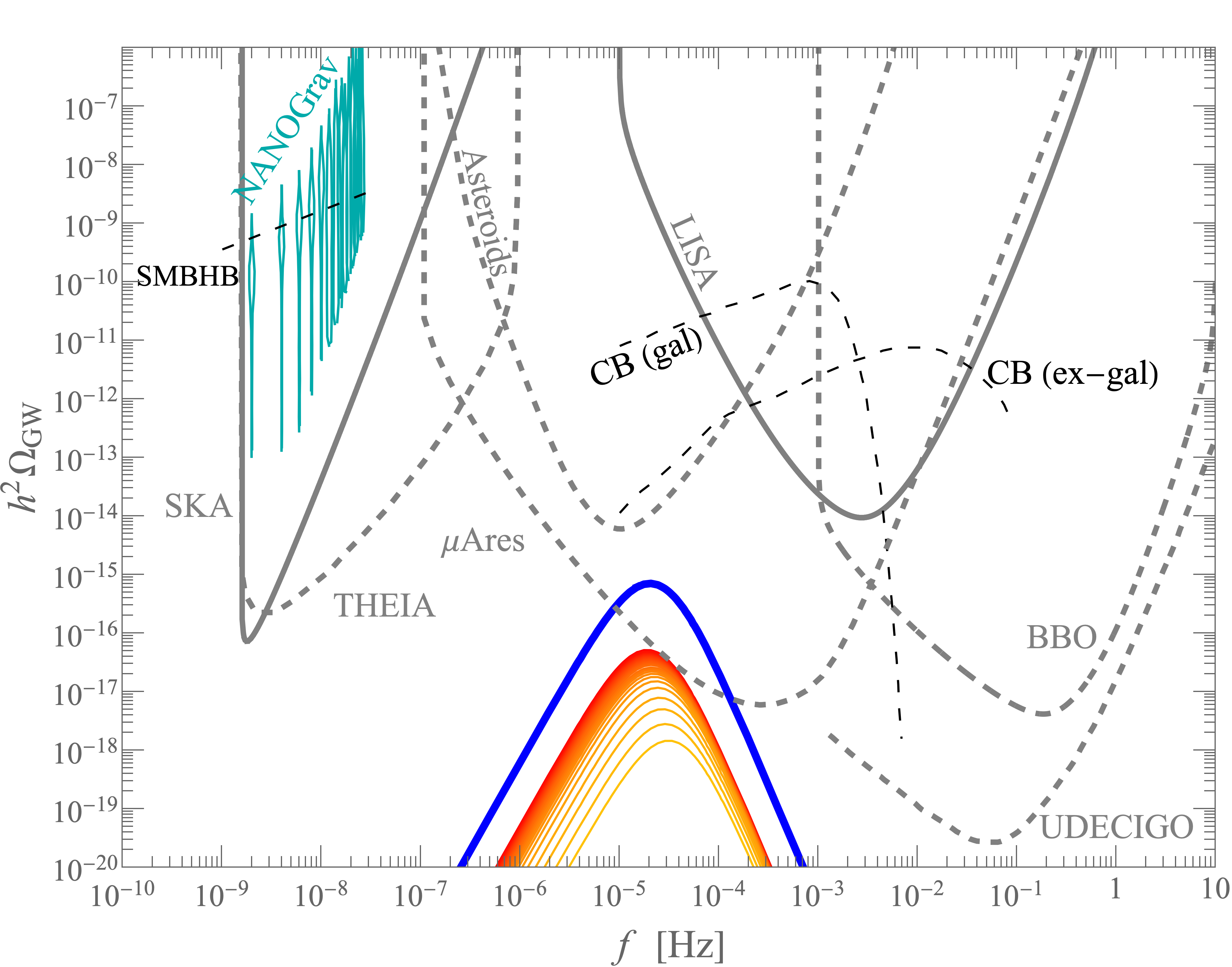} \\
\includegraphics[width=0.48\textwidth]{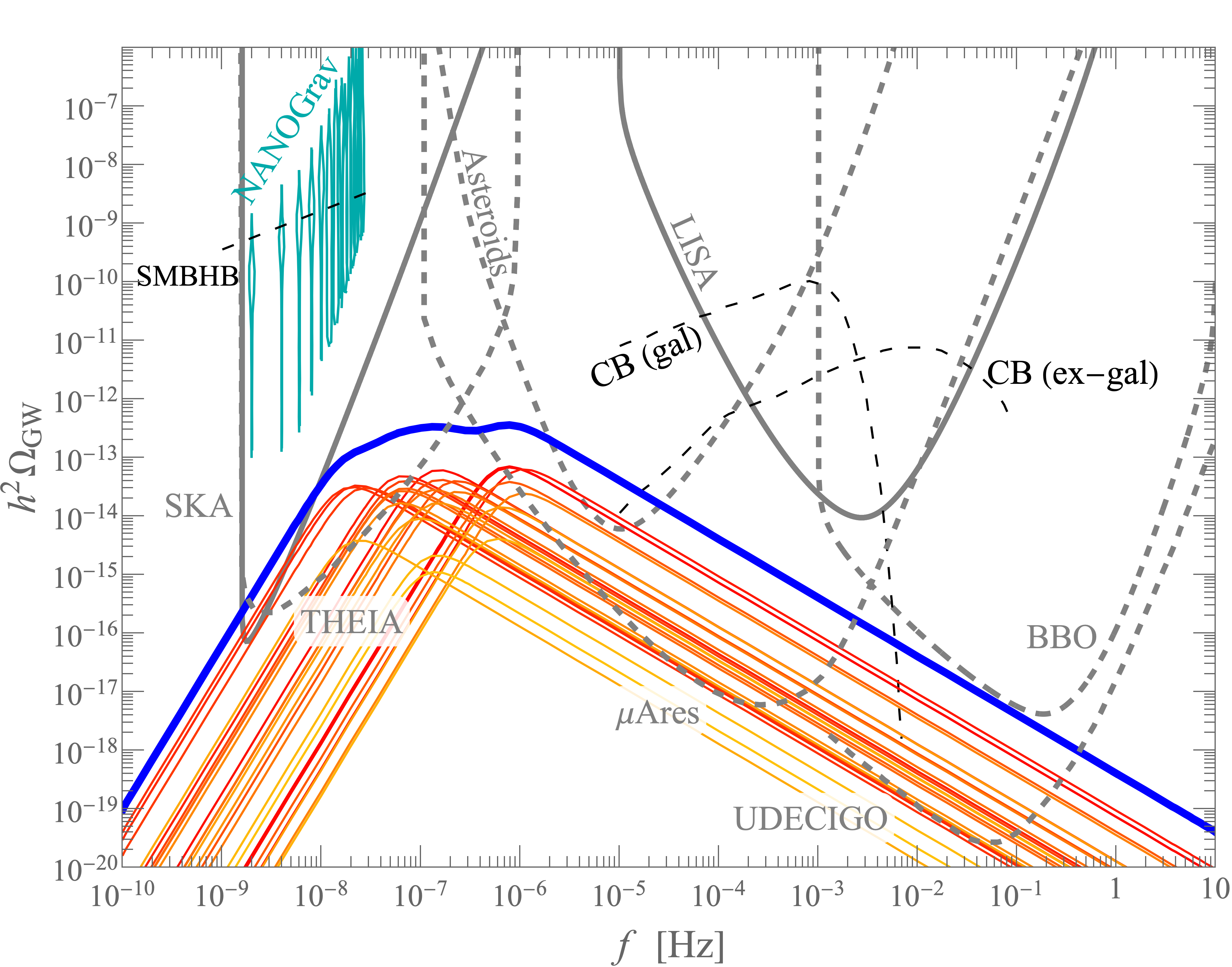} ~
\includegraphics[width=0.48\textwidth]{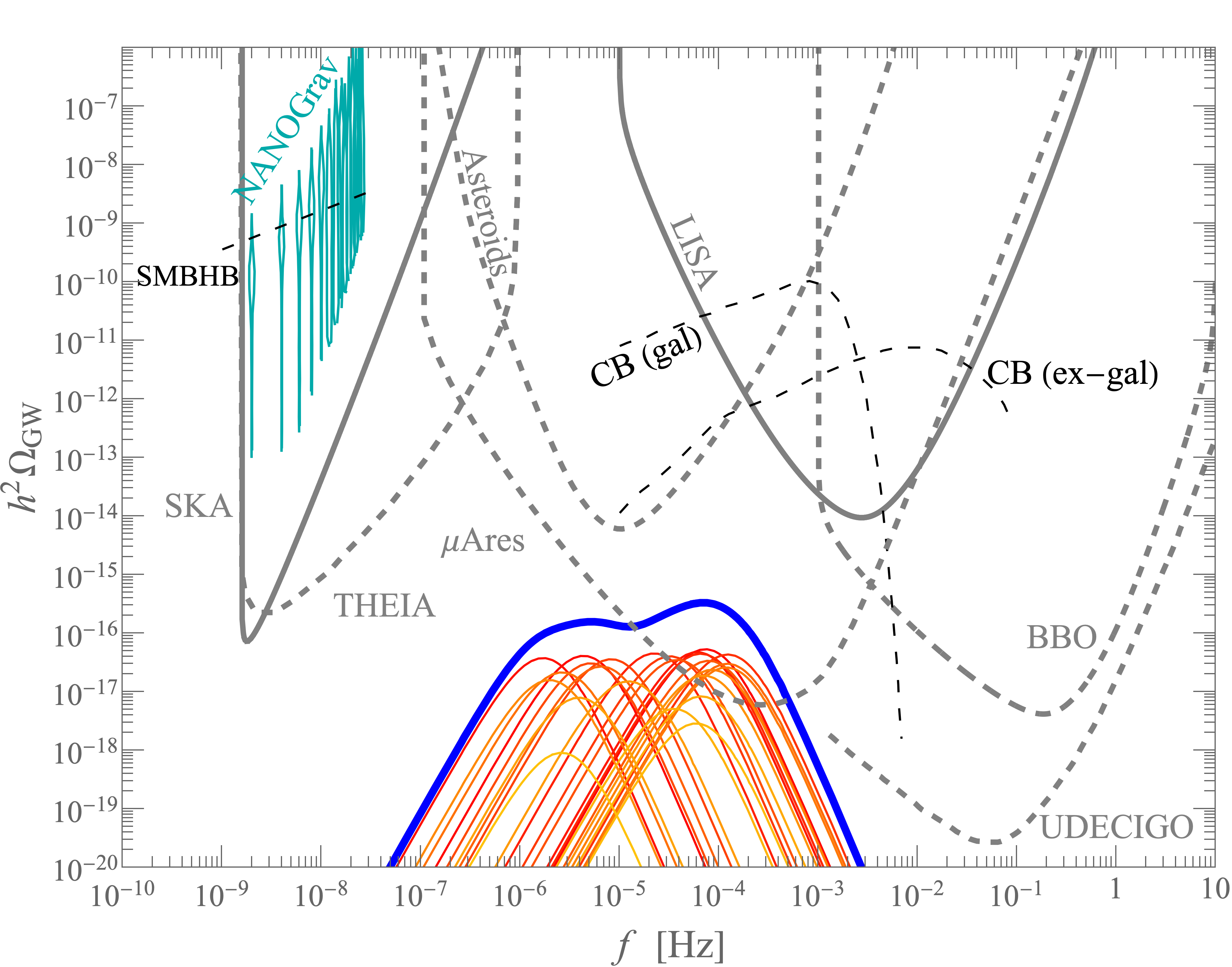}
\caption{
GW spectra from $N$-naturalness for $i_{\rm SM} = 50$.
The figures display the spectra from the 25 lightest exotic sectors, with colors ranging from red to orange to indicate increasingly heavy sectors. 
In each figure, the blue curve denotes the total observed GW spectrum, Eq.~(\ref{eq:GW-total}).
{\it Top left}: GW spectra in the minimal $N$-naturalness model for the runaway scenario, Eq.~\ref{eq:runaway}.
{\it Top right}: Same as top left, but for the non-runaway scenario, Eq.~\ref{eq:nonrunaway}.
{\it Bottom left}: GW spectra in a scenario with non-universal $SU(3)_c$ gauge couplings in each sector for the runaway benchmark, Eq.~\ref{eq:runaway}.
{\it Bottom right}: Same as bottom left, but for the non-runaway scenario.
For the non-universal scenarios, random values of $\delta_i = \delta \alpha_{s_i}/\alpha_s \in \{-0.2,0.2\}$ are chosen, leading to correspondingly different sector confinement scales and peak frequencies according to 
Eqs.~(\ref{eq:modified_lambdaQCDi}) and (\ref{eq:f-peak-0}), respectively. Also shown are sensitivities from various current and future GW observatories; see the main text for further details. 
The Higgs-reheaton mass splitting (\ref{eq:h-phi-splitting}) is fixed to be $\Delta m_{h\phi} = 90$ MeV ($110$ MeV) for the runaway (non-runaway) scenario. Each of these benchmarks predict $\Delta N^{\rm CMB}_{\rm eff} \approx 0.3$. } 
\label{fig:GW-spectra}
\end{figure}

Fig.~\ref{fig:GW-spectra} presents example GW spectra in the minimal $N$-naturalness model for the case $i_{\rm SM} = 50$ for both the runaway (top left) and non-runaway (top right) phase transition scenarios.
We show the spectra from the first 25 exotic sectors, with colors varying from red to orange to denote increasingly heavy sectors. 
The blue line represents the total GW spectra obtained by summing the individual spectra, Eq.~(\ref{eq:GW-total}). 
As can be seen in the figure, the peak frequency varies only slightly from one exotic sector to another, a consequence of the fact that these sectors have nearly identical confinement scales (see Eq.~(\ref{eq:lambdaQCDi})) and thus nearly identical percolation temperatures, $T_i^{{\rm perc},i}$. 
For example, the observed peak frequency for the non-runaway scenario is estimated to be
\begin{align}
\label{eq:f-peak-0}
f_{{\rm p},i}^0 & = 0.53 \left( \beta/H \right)_i  \, v_{{\rm w},i}^{-1} \, H^{\rm perc,i} \left[ \frac{g_{*s,\rm tot}^{0}}{ g_{*s,\rm tot}^{{\rm perc},i}} \right]^{1/3} \frac{T^0}{T^{{\rm perc},i}}  \\
& \approx 20 \, \mu {\rm Hz} \times \left[ \frac{ (\beta/H)_i}{300} \right] \left[ \frac{ 1/\sqrt{3}}{v_{{\rm w},i}} \right]   \left[ \frac{  T_i^{\rm perc,i} }{65 \, {\rm MeV} } \right] \frac{1}{\xi_i^{\rm perc,i}} . \nonumber
\end{align}
Only a small variation across sectors is observed, with the peak frequency increasing as $i$ becomes more negative due to the decreasing temperature ratio $\xi_i^{{\rm perc},i}$ (heavier exotic sectors become increasingly colder). 

Our predicted spectrum is compared to the power law integrated sensitivities (PLIS) representing a signal-to-noise ratio (SNR) threshold of 1 for several future GW observatories in Fig.~\ref{fig:GW-spectra}. These include 
 the SKA PTA~\cite{Janssen:2014dka}; the spaced-based interferometers LISA~\cite{2017arXiv170200786A}, BBO~\cite{Crowder:2005nr}, Ultimate-DECIGO \cite{Braglia:2021fxn}, $\mu$Ares~\cite{Sesana:2019vho}, and asteroid laser ranging~\cite{Fedderke:2021kuy}; and future astrometric measurements \cite{Moore:2017ity} for the proposed THEIA experiment~\cite{2018FrASS...5...11V}. 
Further details on the assumptions underlying our PLIS curves, which are adopted from Refs~\cite{Schmitz:2020syl,Janssen:2014dka,Caprini:2019pxz,Crowder:2005nr,Sesana:2019vho,Fedderke:2021kuy,Garcia-Bellido:2021zgu,Braglia:2021fxn}, can be found in Ref.~\cite{Batell:2023wdb}.
We also display the NANOGrav 15-yr stochastic GW spectrum~\cite{NANOGrav:2023hvm}, along with  estimates for astrophysical foregrounds coming from supermassive black hole binaries (SMBHBs)~\cite{NANOGrav:2023gor} and galactic~\cite{Cornish:2017vip} and extragalactic compact binaries~\cite{Farmer:2003pa}.
We see that SKA, THEIA, $\mu$Ares, asteroid laser ranging, and UDECIGO can potentially probe this $N$-naturalness benchmark under the assumption of a runaway phase transition scenario, Eq~(\ref{eq:runaway}). On the other hand, for a non-runaway scenario (\ref{eq:nonrunaway}), only the proposed $\mu$Ares experiment potentially has sensitivity to test this $N$-naturalness benchmark. 

Although it is intriguing that $N$-naturalness with $i_{\rm SM}>0$ can generate a potentially observable GW signal, the resulting spectrum in the minimal model (Fig.~\ref{fig:GW-spectra}, left panel) closely resembles the standard spectrum characteristic of a single cosmological phase transition, a consequence of the nearly identical sector peak frequencies.
Nevertheless, one can easily envision far more exotic GW spectra in theories with a large number of additional sectors such as $N$-naturalness. 
Here we wish to illustrate this possibility through a slight deformation of the minimal $N$-naturalness model. Specifically, we allow the sectors to exhibit slight variations in their $SU(3)_c$ gauge coupling at the UV cutoff scale $\Lambda_H$, which we parameterize by
\begin{equation}
    \delta_i \equiv \frac{\Delta \alpha_i}{\alpha_{s,{\rm SM}}(\Lambda_H)}.
\end{equation}
Then, the confinement scale $\Lambda_{{\rm QCD},i}$ in the $i$th sector with such a shift is given by 
\begin{equation}
    \Lambda_{{\rm QCD}_i}(\Lambda_H,\delta_i)= 
    \begin{cases}
    \Lambda_H^{1-(1+\delta_i)^{-1}}   (m_t \, m_b \, m_c)^{-\frac{2}{21}(1+\delta_i)^{-1}} (\Lambda_{{\rm QCD}})^{\frac{9}{7}(1+\delta_i)^{-1}} & i<0,\\
    \left[\Lambda_H^{\frac{7}{9}} (m_t \, m_b \, m_c \,)^{\frac{2}{27}} \right]^{1-(1+\delta_i)^{-1}} \displaystyle{\left[\frac{2i+r}{2 i_{\rm SM} +r}\right]^{\frac{1}{9}} }(\Lambda_{{\rm QCD}})^{(1+\delta_i)^{-1}} & i \geq 0 .\\
    \end{cases}
\label{eq:modified_lambdaQCDi}
\end{equation}
Note that in the limit $\delta_{i} \rightarrow 0$, Eq.~(\ref{eq:modified_lambdaQCDi}) reduces to Eq.~(\ref{eq:lambdaQCDi}).
With such shifts $\delta_i$ in the strong gauge coupling, the critical temperatures $T_i^{{\rm crit},i}$ can differ substantially across sectors, producing markedly different GW spectra.
In Fig.~\ref{fig:GW-spectra} we show example GW spectra obtained by choosing random values of this shift in the range  $\delta_i  \in \{-0.2,0.2\}$ for the runaway (bottom left) and non-runaway (bottom right) phase transitions. 
The spectra from individual sectors display clearly separated peak frequencies, while their superposition yields a total GW signal with a pronounced plateau-like structure over a broad frequency range.

Two remarks are in order regarding this scenario. First, as emphasized in Ref.~\cite{Arkani-Hamed:2016rle}, $N$-naturalness does not require the sectors to be identical~\footnote{The assumption of identical sectors (aside from the softly broken permutation symmetry controlling the Higgs mass parameters) in the minimal model is simply a convenient simplification that renders the setup calculable.}, only that our sector is not special in some way. The minor deformation from the minimal model considered here, i.e., varying the strong coupling constant across sectors, is clearly compatible with this criterion and does not impact the naturalness considerations of the scenario.
Second, it is quite natural for such differences in the strong coupling to arise dynamically. For example, suppose each sector contains an additional scalar field with a coupling to the corresponding gluon kinetic term, and whose VEV is drawn from a random distribution. This would generically lead to small, sector-dependent shifts in the strong coupling constant. 

Although we have not attempted to  estimate it here, QCD FOPTs in the lightest SM-like sectors could provide an additional contribution to the stochastic GW background. In these sectors the strange quark is considerably lighter than in the SM, potentially placing them in the region of the QCD phase diagram where a first-order chiral transition occurs (as depicted by the ``Columbia plot''~\cite{Brown:1990ev}). These SM-like sectors would typically exhibit higher critical temperatures, 
shifting the resulting GW peaks to higher frequencies compared to their exotic sector counterparts. Overall, this would add structure to the total GW spectrum, providing an additional observational handle. One important question is the criterion for a FOPT to occur, i.e., how light must the strange quark must be? Ideally this could be explored with lattice methods, but a simple proxy is to require that the kaon of the sector be lighter than its critical temperature, ensuring a light octet of pseudo–Goldstone bosons is present at the onset of the transition (a sign of three light quark flavors). Imposing the condition $m_{K_i} < T_i^{\rm crit,i}$, we find that the $i = 0$
sector can typically undergo a FOPT provided $r$ is not too large, whereas an $i=1$ sector FOPT becomes possible only for relatively large $i_{\rm SM} \gtrsim {\cal O}(100)$.

\subsection{Collider probes of the reheaton}
\label{sec:collider}

The  smoking-gun signature of $N$-naturalness with  $i_{\rm SM} > 0$ is the existence of a new Higgs-mixed scalar—the reheaton—with a mass close to that of the SM Higgs, $m_\phi \approx 125$ GeV.
This near degeneracy with the Higgs is not optional but a mandatory consequence of the resonant reheating mechanism.
In principle, the most direct test of this framework would be to probe the reheaton through its collider phenomenology at the LHC, future Higgs factories, or future high-energy hadron and muon colliders.

A wide variety of effects are conceivable. Direct production of the reheaton, followed by visible decays to SM final states, would yield signatures with the same pattern of relative production modes and branching ratios as the SM Higgs, but with overall rate suppression due to the mixing angle. Invisible reheaton decays into the other sectors could be searched for as missing energy signatures. If measurable, the visible and invisible rates could be correlated with cosmological observables such as $\Delta N_{\rm eff}$, providing a consistency test of the scenario.\footnote{For reference, $\Delta  N_{\rm eff} \approx 0.3$ corresponds to an invisible reheaton branching ratio of about $6\%$.}  The mixing between the reheaton and the SM Higgs induces a universal suppression of the Higgs couplings, which could be probed through precision measurements at Higgs factories as well as high-luminosity/energy hadron or muon colliders. For prompt reheaton decays, the ideal search strategy is to look for two nearby invariant mass peaks and/or distorted line shapes arising from interference between the Higgs and reheaton, especially in clean channels such as $\gamma\gamma$ or $ZZ^{(*)} \rightarrow 4 \ell$~\cite{Dixon:2013haa}.   For sufficiently small mixing angles, the reheaton becomes long-lived, leading to spectacular displaced-vertex signatures at colliders. A measurement of its lifetime would provide a  handle on reheating dynamics, since the lifetime determines the reheating temperature.

However, this appealing phenomenology encounters a severe obstacle: the mixing angle required for successful reheating is extremely small, $\theta_{i_{\rm SM}} \lesssim 10^{-6}$ (see Eq.~(\ref{eq:theta-TRH}) and surrounding discussion). At such tiny mixings, collider production of the reheaton becomes effectively impossible. 
For example, even a high-luminosity 100 TeV hadron collider producing $\sim 10^{10}$  Higgs bosons would yield far fewer than one reheaton produced through Higgs-portal mixing if the reheating bound is respected.  

Here we briefly speculate about two possible avenues that could weaken this conclusion. First, at high temperatures,  the $\phi-h$ mixing angle becomes temperature-dependent and, in particular, may be suppressed compared to its zero-temperature value. Since reheating constraints apply to the high-temperature mixing while collider signals depend on the zero-temperature mixing, it is conceivable that the latter could be much larger than the naive bound. Assessing this possibility requires a dedicated analysis of reheaton cosmological evolution including finite-temperature corrections. We leave this as an interesting direction for future work.

The second possibility concerns production of the reheaton through new UV states. In a UV completion that resolves the full hierarchy problem, it is conceivable that the reheaton may couple to additional heavy states. 
Production of these states
at future energy frontier facilities could lead to enhanced reheaton production, for example, through their cascade decays. While this possibility is highly model-dependent it could circumvent the Higgs-portal suppression.

In summary, even if direct reheaton discovery through Higgs-mixing processes appears unlikely in the minimal $N$-naturalness setup, the framework provides a compelling theoretical motivation for a Higgs-mixed scalar near 125 GeV. Such a state merits close attention at the LHC and future colliders, both because of its intrinsic interest and its possible link to the resonant reheating mechanism of $N$-naturalness.

\section{Outlook}
\label{sec:outlook}

In this work we have introduced $i$-incidental $N$-naturalness, a generalization of the $N$-naturalness framework in which our SM corresponds to one of the heavier sectors with negative Higgs mass-squared, $i_{\rm SM}>0$. We demonstrated that, for a scalar singlet reheaton interacting through a Higgs-portal coupling, resonant reheaton–Higgs mixing generically leads to the preferential reheating of some light SM-like sector, which is then naturally identified with our SM. This provides a distinctive scenario with roughly $i_{\rm SM}$  SM-like and exotic sectors that are lighter than our SM. We examined the bounds on additional relativistic energy densities from new light states in the other sectors ($\Delta N_{\rm eff}$), as well as possible overclosure constraints from heavy stable relics, finding that there is still open parameter space for $i_{\rm SM}$  less than a few thousand.  We further highlighted the possibility of a distinctive stochastic GW background sourced by QCD first-order phase transitions in the numerous lighter exotic sectors. While the presence of a  reheaton that is nearly degenerate with the SM Higgs potentially offers a smoking-gun signature of this scenario, it appears to be challenging to probe this state at the LHC or other future high energy colliders due to its feeble coupling. 

There are several promising directions for future work. 
As in the original $N$-naturalness model, it will be important to study more carefully the cosmological impact of additional stable states in the other sectors. For example, extra neutrinos may behave as warm dark matter and suppress the matter power spectrum~\cite{Bansal:2024afn}. 
It would also be valuable to examine how robust the resonant reheating mechanism is under different assumptions regarding the statistical distributions of Higgs mass-squared parameters, rather than the even spacing assumed in Eq.~(\ref{eq:mH-distribution-1}).
Moreover, a more refined treatment of reheating, including finite-temperature corrections and relaxing the instantaneous-reheating approximation, could lead to a more accurate determination of the reheating temperature and the maximal reheaton–Higgs mixing. Finally, the basic mechanism of populating the SM via resonant mixing opens up new model-building possibilities for the reheaton sector and its interactions, which merit further exploration.

\section{Acknowledgements}
We thank Nima Arkani-Hamed, Pouya Asadi, Tao Han, and Sunghoon Jung for helpful discussions. 
The work of B.B., W.H., and M.L. is supported by the U.S. Department of Energy under grant No. DE–SC0007914.  
M.L. is also supported by the National Science Foundation under grant No. PHY-2112829.
The work of A.G. is supported by the GRASP initiative at Harvard University.

\appendix

\section{Sums}
\label{sec:sums}

In this appendix, we provide some details on performing the sums over sectors needed to obtain the approximate formulae for  $\Delta N_{\rm eff}$ presented in Sec.~\ref{sec:DNeff}.  
We begin with the contribution from light exotic sectors, Eq.~(\ref{eq:DNeff-exotic}). We require the sum
\begin{align}
\label{eq:DNeff-sum-exotic}
\sum_i \Gamma_{\phi \rightarrow H_i H_i^\dag} & \approx \int d  i  \, \Gamma_{\phi \rightarrow H_i H_i^\dag}   = \frac{2 i_{\rm SM}+ r}{m_h^2 }\frac{a^2}{8 \pi m_\phi} \int_0^{m_\phi^2/4} dm^2_{H_{i}} \sqrt{1-\frac{4m_{H_{i}}^2}{m_\phi^2}}  \\
& = \frac{a^2 m_\phi}{48  \pi m_h^2} (2 i_{\rm SM}+r). \nonumber
\end{align}
We have included the dominant contribution from exotic sectors receiving on-shell two-body reheaton decays, Eq.~(\ref{eq:Gamma-HH}).
In the second step above, the first prefactor arises from the change of measure when transforming the integration variable, $m^2_{H_i} = (m_h^2/2)(2i+r)/(2 i_{\rm SM}+r)$. 
Using Eq.~(\ref{eq:DNeff-sum-exotic}), the reheaton decay width to the SM, $\Gamma_{\rm SM} \approx \theta^2_{i_{\rm SM}} \Gamma_h$, and working to leading order in $\Delta m_{h\phi} \equiv m_h - m_\phi$, we readily obtain Eq.~(\ref{eq:DNeff-exotic}). 

A similar exercise can be carried out for the light SM-like sectors, with the modification that we must consider reheaton decays to pairs of on-shell $W^\pm$, $Z$, and Higgs bosons, 
\begin{align}
\label{eq:DNeff-sum-SML}
\sum_i \left( \Gamma_{\phi \rightarrow W^+_i W^-_i}+\Gamma_{\phi \rightarrow Z_i Z_i}+\Gamma_{\phi \rightarrow h_i h_i} \right)  
& \approx  \frac{1}{8} 
\left[2 \, I\!\left(\frac{m_W^2}{m_h^2}\right)+  I\! \left(\frac{m_Z^2}{m_h^2} \right) + 1 \right] \sum_i \Gamma_{\phi \rightarrow H_i H_i^\dag} \\
& \approx  1.33 \sum_i \Gamma_{\phi \rightarrow H_i H_i^\dag}, \nonumber
\end{align}
leading to the result in the main text. 
Here the integral $I(\hat x)$ is given by
\begin{equation}
I(\hat x) \equiv 24 \hat x \int_0^1 dx (x - 4 \hat x)^{-2} (1-x)^{1/2} (1-x+\tfrac{3}{4} x).
\end{equation}

Finally, for the SM-like sectors in the resonant region we have 
\begin{align}
\label{eq:DNeff-sum-Res}
\sum_{i \in i_{\rm Res}} \frac{\Gamma_i}{\Gamma_{\rm SM}} & \approx (2 i_{\rm SM}+r )\frac{(m_h^2 - m_\phi^2)^2}{2 m_h^4 } \int_{i \in i_{\rm Res}} dm^2_{h_i} \frac{m_{h_i}^2 }{(m_{h_i}^2 - m_\phi^2)^2 } R_{\Gamma_{h_i}} \\
& = (2 i_{\rm SM}+r )\frac{(m_h^2 - m_\phi^2)^2}{2 m_h^4 } \int_{i \in i_{\rm Res}}  dx  \frac{x}{(x-1)^2} R_{\Gamma_{h_i}}.
\end{align}
where in the second step we have changed integration variables via $x  = m^2_{h_i}/m_\phi^2$. 
The width ratio $R_{\Gamma_{h_i}}$, defined in Eq.~(\ref{eq:width-ratio}), is a slowly varying function of $i$. Since the integral has support in the resonance region, we can approximate $R_{\Gamma_{h_i}} \approx R_{\Gamma_{h_i}}\big\vert_{x = 1} \approx 1$. 
To include the nearby SM-like sectors undergoing resonant $\phi-h_i$ mixing, we break up the integral into two pieces by excluding a small region of size $\Delta x = 2 \delta$ around the resonance,
\begin{align}
\label{eq:int-Res}
 \int_{i \in i_{\rm Res}}  dx  \frac{x}{(x-1)^2}  & \approx  \left[ \int_{x_{\rm min}}^{1-\delta} + \int_{1+\delta}^{x_{\rm max}}  \right]  \, dx \,   \frac{x}{(x-1)^2} \\
 &  = [J(1-\delta)- J(x_{\rm min}) ] +  [J(x_{\rm max})- J(1+ \delta) ]  \approx \frac{2}{\delta}, \nonumber
\end{align}
where $J(x) = (1-x)^{-1} + \ln(x-1)$.
In the last step, we made use of the fact that the contributions far away from the resonance region, $J(x_{\rm min})$, $J(x_{\rm max})$,  are small and can be neglected, and we also took the limit $\delta \ll 1$. 
The excluded region is defined by the spacing of Higgs masses near resonance, 
\begin{align}
\label{eq:delta-res}
\delta \sim \frac{m^2_{h_{i_{\rm SM}}}}{m^2_\phi} - \frac{m^2_{h_{i_{\rm SM}-1}}}{m^2_\phi} = \frac{m^2_h}{m^2_\phi} \frac{2}{2 i_{\rm SM}+ r} \approx  \frac{2}{2 i_{\rm SM}+ r}.
\end{align}
Taking Eqs.~(\ref{eq:DNeff-sum-Res},\ref{eq:int-Res},\ref{eq:delta-res}), we obtain Eq.~(\ref{eq:DNeff-Res}) in the main text. 

In the large $i_{\rm SM}$ regime, we must account for the Higgs width given the narrow spacing of Higgs masses in nearby sectors. 
The contributions from sectors in the resonance region dominates $\Delta N_{\rm eff}$, and Eq.~(\ref{eq:DNeff-sum-Res}) is modified to the form
\begin{align}
\label{eq:DNeff-sum-Res-1}
\sum_{i \in i_{\rm Res}} \frac{\Gamma_i}{\Gamma_{\rm SM}} & \approx (2 i_{\rm SM}+r )\frac{(m_h^2 - m_\phi^2)^2+m_h^2 \Gamma_h^2}{2 m_h^4 } \int_{i \in i_{\rm Res}} dm^2_{h_i} \frac{m_{h_i}^2 }{(m_{h_i}^2 - m_\phi^2)^2 + m_{h_i}^2 \Gamma_{h_i}^2 } R_{\Gamma_{h_i}}.
\end{align}
Due to the narrow Higgs width, the denominator in the integral is well approximated by a delta function, and considering the resonance region $m_\phi \sim m_h$ and the large $i_{\rm SM}$ limit, we obtain
\begin{align}
\label{eq:DNeff-sum-Res-2}
\sum_{i \in i_{\rm Res}} \frac{\Gamma_i}{\Gamma_{\rm SM}} & \approx  i_{\rm SM} \frac{\Gamma_h^2}{m_h^2} \frac{\pi}{m_h \, \Gamma_h} \int dm^2_{h_i} 
\delta(m^2_{h_i} - m_h^2) m_{h_i}^2  R_{\Gamma_{h_i}}  =  \frac{\pi \Gamma_h }{m_h }  i_{\rm SM}. 
\end{align}
Finally, since we have integrated over the entire resonance region, we should subtract off the SM contribution when computing $\Delta N_{\rm eff}$. Doing so yields  Eq.~(\ref{eq:DNeff-Res-1}).

\bibliographystyle{apsrev4-1}
\bibliography{refs}
\end{document}